\documentclass[preprint,12pt]{elsarticle}

\usepackage[T1]{fontenc} 
\usepackage[utf8]{inputenc}
\usepackage{multirow,subcaption}
\usepackage{amsmath, amssymb}
\usepackage{caption, subcaption}
\usepackage{multirow}
\usepackage{color}

\begin{document}

\begin{frontmatter}
\title{Ho\v{r}ava-Lifshitz cosmology in light of new data}
\author[a]{Nils A. Nilsson}
\ead{albin.nilsson@ncbj.gov.pl}
\author[a]{Ewa Czuchry}
\ead{ewa.czuchry@ncbj.gov.pl}
\address[a]{National Centre for Nuclear Research, Ho\.za 69, 00-681, Warsaw, Poland}

\begin{abstract}
We present new observational constraints on Lorentz violating Ho\v{r}ava-Lifshitz cosmological scenarios using an updated cosmological data set from Cosmic Microwave Background (Planck CMB), expansion rates of elliptical and lenticular galaxies, JLA compilation (Joint Light-Curve Analysis) data for Type Ia supernovae (SneIa),  Baryon Acoustic Oscillations  (BAO) and priors on the Hubble parameter with an alternative parametrisation of the equations. Unlike in other approaches we
    consider the curvature parameter $\Omega_k$ as a free  parameter in the analysis we considered the parameters $\Omega_k$ and $\Delta N_\nu$ as completely free, which helped to place new, updated bounds on several of the theory parameters.  Remarkably, the detailed balance scenario exhibits positive spatial curvature to more than 3$\sigma$, whereas for further theory generalizations we found evidence for positive spatial curvature at 1$\sigma$. This could create circumstantial evidence from
    observations and could be used to single out distinct formulations and scenarios.
\end{abstract}

\end{frontmatter}


\section{Introduction}
One of many challenging problems in modern theoretical physics is how to merge quantum field theory with general relativity. Attempts at formulating a quantum theory for gravitation have been made for many decades, but there is still no theory which clearly stands out from its competitors. One of the main obstacles one encounters when trying to quantise general relativity using standard techniques is that it is not a perturbatively renormalisable theory. This is indeed a serious problem as general relativity breaks down at small scales, and alternative formulations for a quantum theory of gravity have been developed in order to deal with this issue. For example, string theory and loop quantum gravity have been developed for this reason, and although these theories resolve some problems (such as singularities) there are few phenomenological channels through which to test these theories~\cite{Quevedo:2016tbh,Girelli:2012ju}. Although no theory of quantum gravity is known it is possible to predict some of its features. For example, it is likely to contain the three constants $c$ (speed of light, relativity), $G$ (gravitational constant, gravitation), and $\hbar$ (Planck's constant, quantum mechanics). Using these constants, we can construct a characteristic energy scale for quantum gravity as $E_{Pl} = \sqrt{\hbar c^5/G}$. This is to be interpreted at the ``frontier`` beyond which we can expect quantum gravity effects to manifest themselves. If this is indeed the case, we can only expect to find quantum gravity effects in regions with very high energy, strong gravity, and high velocities. The phenomenological avenues to probe these regions are sparse, and do not in general allow themselves to be recreated in the  laboratory, so we are forced to rely on astrophysical observations. One possibility is that of TeV particle astrophysics, as observed with the next generation of telescopes. There are several scenarios in which quantum gravity effects could affect the opacity of the Universe to such high-energy particles~\cite{Kifune:1999ex, Protheroe:2000hp}. These effects would multiply with the distance, and could in principle be detectable with (for example) the CTA~\cite{Fairbairn:2014kda}.

An area of observation which has recently opened up a new window to the Universe is that of \emph{gravitational wave astronomy}. Recently, LIGO found evidence of a binary black hole merger~\cite{Abbott:2016blz}, which was consistent with the prediction from general relativity for such events. Moreover, LIGO and VIRGO have also observed a binary neutron star merger~\cite{TheLIGOScientific:2017qsa} which was accompanied by a short $\gamma$-ray burst picked up by the \textsc{FERMI} and
\textsc{INTEGRAL} $\gamma$-ray telescopes~\cite{Troja:2017nqp}. This put strong constraints on the speed of tensorial gravitational waves, as the difference $\Delta t$ from the speed of light $c$ was found to be $-3\cdot 10^{-15}<\Delta t/c<7\cdot 10^{-16}$~\cite{Monitor:2017mdv}. As opposed to general relativity, which permits only tensor metric perturbations, many alternative models of gravity have more degrees of freedom. For example, scalar-tensor theories such as Brans-Dicke theory permits
a transverse-scalar mode, and Ho\v{r}ava-Lifshitz gravity allows both vector and tensor modes~\cite{PhysRev.124.925, Blas:2011zd}. This has also been discussed in the context of Lorentz-violating theories, for example in~\cite{Gumrukcuoglu:2017ijh} (Ho\v{r}ava gravity) and in~\cite{Oost:2018tcv} (Einstein-\AE ther theory). In general, gravitational wave data places strong constraints on the coupling constants of the theories, as well as the speed of the tensor mode. Moreover,
these observations also place constraints on the Lorentz breaking scale (often denoted $M_*$) in Ho\v{r}ava gravity~\cite{Gumrukcuoglu:2017ijh}.

Recently, the LIGO and VIRGO team reported the first ever direct limits of the strain of scalar and vector modes at $< 1.5 \cdot 10^{-26}$ at $95\%$~\cite{Abbott:2017tlp, Abbott:2018utx}. Prior to this, the magnitude of scalar modes were completely unconstrained, and this new discovery can place tight bounds on alternative theories of gravity, such as Ho\v{r}ava-Lifshitz. This new field of multimessenger astronomy has the potential to teach us more about strong gravity events and will be some of the most important ways of probing new physics in the future.

The fact that general relativity has been tested and found consistent with virtually all observations tells us that it is a very good model for the IR behaviour of quantum gravity. Thus, any candidate quantum gravity model proposed must have general relativity as its low energy limit. It therefore makes sense to look at proposals for UV-completions of general relativity, for example~\cite{ArkaniHamed:1998rs, Dvali:2007hz}. In general these include a cutoff energy scale beyond which general
relativity breaks down. If this energy scale is low enough (in the TeV range) there may be a lot of accessible phenomenology in these models. Another interesting proposal in this category is Ho\v{r}ava gravity, which is a concrete proposal of a UV complete theory of gravity, and possibly of quantum gravity~\cite{Horava:2009uw,Horava:2010zj}. It is also referred to as Ho\v{r}ava-Lifshitz (HL) gravity since if contains a fixed point in the UV with an anisotropic Lifshitz scaling between space and
time, explicity breaking Lorentz invariance in the UV. A lot of work has gone into this model, including studies on cosmological solutions~\cite{Mukohyama:2010xz,Kiritsis:2009sh,Calcagni:2009ar}, dark radiation and braneworlds~\cite{Calcagni:2009ar, Saridakis:2009bv}, and the resolution of the initial singularity~\cite{Brandenberger:2009yt, Czuchry:2010vx, Czuchry:2009hz}. Several papers have placed bounds on different regions of the Ho\v{r}ava-Lifshitz framework, for example using cosmological
data~\cite{Dutta:2009jn}, binary pulsars~\cite{Yagi:2013ava,Yagi:2013qpa}, and in the context of dark energy~\cite{Park:2009zr}. This has also been    done~\cite{Frusciante:2015maa} in the effective field theory formalism of the the extended version  of   Ho\v{r}ava theory~\cite{Blas:2009qj}. Other authors have placed bounds on more general Lorentz violation in the context of dark matter and dark energy, for example~\cite{Audren:2014hza, Audren:2013dwa}.

The breaking of Lorentz invariance mentioned above may seem counterintuitive, as it is one of the fundamental principles of modern physics and is strongly supported by experiment. In fact, every single experiment so fas been consistent with Lorentz invariance~\cite{Coleman:1998ti, Kostelecky:2008ts}. This means that breaking Lorentz symmetry has to be done with great care. However, there is \emph{no a priori reason} to expect that a theory of quantum gravity should uphold Lorentz invariance, as according to our understanding, spacetime takes on a quantum nature in the Planck regime, and as we have mentioned before, continuous classical spacetime emerges as a low energy limit of this quantum theory. Then, as Lorentz symmetry is a \emph{continuous symmetry} of nature, it bears to reason that it might not exist at all in the quantum regime, but rather emerge as an IR symmetry of nature. Once Lorentz symmetry is no longer a concern, one may include higher-order derivatives in the Lagrangian in order to cure divergencies in the UV~\cite{Wang:2017brl}.

There is an open discussion about instabilities and pathologies in different formulations of HL-type theories (see e.g.  \cite{Sotiriou:2009bx,Blas:2009qj, Sotiriou:2010wn}  or a recent review \cite{Wang:2017brl}).
The original HL theory suffers from many inconsistencies and shortcomings, the most significant of which being the existence of a parity violating term \cite{Sotiriou:2009bx},
problems with ghost instabilities and strong coupling at very low energies \cite{Blas:2009yd, Charmousis:2009tc}, wrong sign and very large value of cosmological constant \cite{Vernieri:2011aa, Appignani:2009dy}, problems with power counting renormalisation of the scalar mode propagation \cite{Vernieri:2015uma,Colombo:2015yha}.  Some of those issues (ghost instabilities and the sign of the cosmological constant) are discussed in more detail later on in this paper and we discuss alternative
solutions to them, like carrying out an analytic continuation of the constant parameters (\cite{Lu:2009em}). Other problems arise from the conditions imposed in the original Ho\v{r}ava formulation of the theory; the detailed balance condition, where it is assumed that potential part of the action is given from the superpotential, thus limiting the number of its terms and independent couplings. There is also the projectability condition which assumes that the lapse function $\cal N$ depends only
on time ${\cal N}={\cal N}(t)$.  Some authors (\cite{Charmousis:2009tc},  end note) claim that the non-projectable version leads to serious strong coupling problem and does not have a viable GR limit at any energy scale. On the other hand  more recent considerations, e.g.  \cite{Pospelov:2010mp, Vernieri:2011aa}, claim the opposite. These authors state that most of the  shortcomings (like parity violation, problems with renormalisation of the scalar mode and its IR behaviour) can
be cured by adding additional terms to the superpotential and relaxing projectability condition, while still keeping detailed balance (or eventually softly breaking it). However later works provide evidence for problems with power-counting renormalizability of the scalar mode (spin-0 graviton) in the detailed balance version of the theory \cite{Vernieri:2015uma}. Recent formulation of the so called "mixed-derivative" Ho\v{r}ava gravity \cite{Colombo:2015yha,Coates:2016zvg} claim that this formulation is actually power counting renormalizable.

There is still an open issue within HL-type theories, namely the large value of the cosmological constant, as there is a descrepancy of at least 60 orders of magnitude between the value demanded by detailed  balance \cite{Vernieri:2011aa} and the observed one. It is discussed in \cite{Appignani:2009dy} that adding the effects of very large vacuum energy of quantum matter fields may cancel out the negative sign and magnitude of the bare cosmological constant leaving the tiny observed
residual. There is also  problem of a huge predicted value for the quantum vacuum (review in \cite{Carroll2001}), as there is a large difference between the magnitude of the vacuum energy expected from zero-point fluctuations and scalar potentials and the observed value, and maybe those issues have common solution.

Taking into account various problems and contradicting statements in the works studying different HL-type theories from the analytical side, it is worth to investigate how different formulations fit the observational data.  Currently Ho\v{r}ava gravity and its extensions are not  ruled out  observationally (although the recent binary neutron star merger GW170817 \cite{Monitor:2017mdv} provides tight bounds on some parameters \cite{Gumrukcuoglu:2017ijh}); thus further observational constraints should either rule out its specific scenarios or the whole model, or in case of agreement with observations provide a better justification for deeper theoretical research. It is still  believed  that HL theory might provide a promising cosmological scenario solving some shortcomings of classical GR, like non-renormalisability. We are actually interested in fitting cosmological scenarios predicted by different version of HL theories to existing observational data.  We believe that it is a separate issue from theoretical analysis, as observational constraints may provide valuable input for theoretical analysis.

Studies providing bounds on HL theory constants and parameters based on observational data already exist in the literature. Some are based on the two simplest HL scenarios and quite old cosmological data, like \cite{Dutta:2009jn} (non-projectable version with and without detailed balance). Others like \cite{Frusciante:2015maa} fit data to the more general  extension of Horava theory \cite{Blas:2009qj}, and also consider linear perturbations around that background. However, this research
  is based on a flat model (in our case the curvature density parameter is left as a free parameter). As a part of a bigger project we would like to place new bounds on parameters appearing in Ho\v{r}ava-Lifshitz cosmology based on the more recent data. As a first step we would like to fit  the data to the two simplest cosmological models. We will look at scenarios based on the original Ho\v{r}ava formulation \cite{Horava:2009uw,Horava:2010zj} with detailed balance condition imposed, as well as one with this
  condition relaxed and with the generalised form of the gravitational action (Sotiriou-Visser-Weinfurtner (SVW) generalisation)~\cite{Sotiriou:2009bx}. This way we can see the effect of applying newer observational data and different parameterisation  on the values of HL cosmological parameters. In the next step we are going to include bigger sets of data, like PANTHEON SN1a catalogue \cite{Scolnic:2017caz} and  gamma-ray bursts \cite{Liu:2014vda} and/or apply it to a more general HL formulation, like \cite{Blas:2009qj} - the so called "healthy" or "consistent" extension of the original formulation. There are works \cite{Frusciante:2015maa}  putting constraints on this extension written in the effective field theory framework, but the latter is based on a flat background, limiting the number of parameters.
In this work we have focused on a simpler cosmological model and a smaller data set in order to better capture the influence of new data on the model as compared to earlier works. We also used the least number of free parameters and more natural parameterisation of the other ones.  By increasing the size of the data sets incrementally we gain a better understanding as to the sensitivity of the model to data with higher resolution and range.

In this paper, being the first one in a planned series,
we provide improved observational constraints on two basic HL scenarios, which are based on the most recent cosmological data set from
Cosmic Microwave Background (Planck CMB)~\cite{planck2015}, expansion rates of elliptical and lenticular galaxies~\cite{Moresco15}, JLA compilation (Joint Light-Curve Analysis) data for Type Ia supernovae (SneIa)~\cite{JLA},  Baryon Acoustic Oscillations  (BAO)~\cite{WiggleZ,SDSS12,Lyman} and priors on the Hubble parameter~\cite{Bennett14}. These updated observational results, together with an alternative parameterisation of the Friedmann equations, provided much tighter constraints on parameters of the HL-type theories, more prone to further observational verification.

The paper is organised as follows: in Section 2 we go through brief review of Ho\v{r}ava-Lifshitz gravity and describe setup for observational constraints in two cases, the original one and the SVW generalisation~\cite{Sotiriou:2009bx}.   In Section 3 we present the new bounds on the theory parameters, based on a large updated cosmological dataset, and we conclude  in Section 4. The details of our numerical analysis are presented in the Appendix.

\section{Ho\v{r}ava-Lifshitz Cosmology}

\subsection{Basics}
As we mentioned previously, one of the main obstacles on the road to quantum gravity is the fact that general relativity is non-renormalisable. This is easy to see if we notice that the expansion of any quantity $\mathcal{F}$ in terms of the gravitational constant can be written as~\cite{Wang:2017brl}:
\begin{equation}
\mathcal{F} = \sum_{n=0}^\infty  a_n\left(G_N E^2\right)^n,
\end{equation}
where $E$ is the energy of the system, $a_n$ is a numerical coefficient and $G_N$ is the gravitational coupling constant.
From this we see that then $E^2 \geq G^{-1}$, this expansion diverges, and it is expected that general relativity loses perturbative renormalisability at such high energies.

It is possible to improve the ultraviolet behaviour of general relativity by introducing higher-order derivative terms to the Einstein-Hilbert Lagrangian. For example, we can write it as: $S = \int$ d$^4x \sqrt{-g}(R+R_{\mu\nu}R^{\mu\nu})$, which changes the graviton propagator from $1/k^2$ into $1/(k^2-G_Nk^4)$. The new term proportional to $k^{-4}$ will indeed cancel the ultraviolet divergence, but the theory is now non-unitary and equipped with a massive spin-2 ghost~\cite{Wang:2017brl}. In fact, this ghost comes from the fact that this higher-order gravity theory has time derivatives of $\mathcal{O}>2$. In the example above, the resulting field equations are fourth order. In order to address this problem, Ho\v{r}ava decided to evade the ghost by constructing a higher-order theory of gravity where only the \emph{spatial} derivatives are of $\mathcal{O}>2$, while keeping second-order time derivatives only~\cite{Horava:2009uw}. This requires an explicit breaking of Lorentz invariance in the ultraviolet, but to be consistent with all current experiments, which have so far failed to detect any significant signals of Lorentz invariance violation, this symmetry has to be \emph{restored} in the infrared limit. The way Ho\v{r}ava overcame this hurdle was to introduce an anisotropic scaling of space and time in the ultraviolet (also known as Lifshitz scaling). This scaling takes the form (in a 4-dimensional spacetime):
\begin{equation}
t \to b^{-z}t, \, x^i \to b^{-1}x^i, \, (i=1,2,3),
\end{equation}
where $z$ is a critical exponent. To restore Lorentz invariance we need to set $z=1$, but to have power-counting renormalisability we need to have $z\geq 4$~\cite{Wang:2017brl}. The most common choice is to set $z=3$. Thus, ultraviolet Lorentz violation lies at the core of Ho\v{r}ava-Lifshitz gravity and cosmology. Ho\v{r}ava assumed that it is broken down to $t \to \xi_0(t), \, x^i \to \xi^i(t,x^k)$, which preserves the spatial diffeomorphisms. This theory acquires a symmetry group denoted Diff[M,$\mathcal{F}$], which is known as foliation-preserving diffeomorphisms~\cite{Horava:2009uw,Wang:2017brl}. This allows for arbitrary changes of the spatial coordinates on each constant time slice, and also picks out a preferred time foliation of spacetime, which breaks Lorentz invariance.

Because of the anisotropic scaling between space and time, it is convenient to introduce the ADM decomposition of the spacetime metric in a preferred foliation:
\begin{equation}
\text{d}s^2 = -N^2\text{d}t^2 + g_{ij}(\text{d}x^i+N^i\text{d}t)(\text{d}x^j+N^j\text{d}t),
\end{equation}
where the dynamical variables of the theory now are the lapse function $N$, the shift vector $N^i$, and the spatial metric $g_{ij}$ ($i$, $j=1,2,3$). Given this, we can write down the most general action for the theory as:
\begin{equation}\label{eq:gen_action}
S = \int \text{d}^3x\text{d}tN\sqrt{g}\left[K^{ij}K_{ij}-\lambda K^2-\mathcal{V}(g_{ij})\right],
\end{equation}
where $g$ is the determinant of the spatial metric $g_{ij}$, $\lambda$ is a running coupling (dimensionless), and $\mathcal{V}$ is a potential term. Moreover, $K_{ij}$ is the extrinsic curvature:
\begin{equation}
K_{ij} = \frac{1}{2N}\left(\dot{g}_{ij}-\nabla_iN_j-\nabla_jN_i\right),
\end{equation}
where an overdot denotes a full derivative with respect to the time $t$. The trace of $K_{ij}$ is denoted $K$. The potential $\mathcal{V}$ depends only on the spatial metric and its spatial derivatives, and is also invariant under three-dimensional diffeomorphisms~\cite{Blas:2009qj}. It contains \emph{only} operators of dimension 4 and 6 which can be constructed from the spatial metric $g_{ij}$.

\subsection{Detailed Balance}
A way to simplify the action (\ref{eq:gen_action}) is to impose the so-called detailed balance condition, in which we assume that it should be possible to derive $\mathcal{V}$ from a superpotential $W$~\cite{Horava:2009uw,Sotiriou:2010wn,Vernieri:2011aa}:
\begin{equation}
\mathcal{V} = E^{ij}\mathcal{G}_{ijkl}E^{kl}, \quad E^{ij} = \frac{1}{\sqrt{g}}\frac{\delta W}{\delta g_{ij}},
\end{equation}
and
\begin{equation}
\mathcal{G}^{ijkl} = \frac{1}{2}\left(g^{ik}g^{jl}+g^{il}g^{jk}\right) - \lambda g^{ij}g^{kl}.
\end{equation}
From this requirement, the most general action for Ho\v{r}ava-Lifshitz gravity is given by~\cite{Sotiriou:2010wn}:
\begin{equation}\label{eq:action}
\begin{aligned}
S_{db} = \int \text{d}t\, \text{d}^3x\sqrt{g}N&\Bigg[\frac{2}{\kappa^2}\left(K_{ij}K^{ij}-\lambda K^2\right)+\frac{\kappa^2}{2\omega^4}C_{ij}C^{ij}-\frac{\kappa^2\mu}{2\omega^2}\frac{\epsilon^{ijk}}{\sqrt{g}}R_{il}\nabla_jR^l_k \\&+\frac{\kappa^2\mu^2}{8}R_{ij}R^{ij}+\frac{\kappa^2\mu^2}{8(1-3\lambda)}\left(\frac{1-4\lambda}{4}R^2+\Lambda R -3\Lambda^2\right)\Bigg],
\end{aligned}
\end{equation}
where
\begin{equation}
C^{ij}=\epsilon^{ikl} \nabla_k \left (R^j_{\ l}-\frac{1}{4}R\delta^j_l\right)
\end{equation}
is the Cotton tensor, $\epsilon^{ikl}$ is the totally antisymmetric tensor, and the parameters $\kappa, \omega$, and $\mu$ have mass dimension $-1,0$, and $1$, respectively. The action \eqref{eq:action} for HL gravity has been obtained from the original  one \cite{Horava:2009uw}  by carrying out    an analytic  continuation (e.g. \cite{Lu:2009em}) of constant parameters: $\mu \mapsto i\mu$ and $\omega^2\mapsto-i\omega^2$. This enabled to obtain positive values of  the cosmological constant $\Lambda$ (that correspond to current observational results) in the low energy limit of the theory, unlike in the original formulation.

The coupling constant $\lambda$ is dimensionless. In general, it runs  (logarithmically in the high energy limit --  in UV)  and may eventually reach one the three infrared (IR) fixed points (\cite{Horava:2009uw}): $\lambda=1/3$, $\lambda=1$ or $\lambda=\infty$. The range  $1>\lambda>1/3$ leads to ghost  instabilities in the IR limit of the theory  \cite{0264-9381-27-7-075005}.   However, this range of $\lambda$ is exactly the flow-interval between the UV and IR regimes. The only physically interesting case that remains,
allowing for a possible flow towards GR -- at $\lambda=1$ --  is the regime $\lambda\ge1$. Region $\lambda\le 1/3$ is disconnected from $\lambda=1$ and cannot be included in realistic considerations.

 We expect that the action \eqref{eq:action} corresponds to the Einstein-Hilbert near the IR limit of the theory.
This happens for the speed of light $c$ and gravitational constant $G$ expressed in terms of HL parameters as follows:
\begin{equation}\label{eq:cosmoconstants}
 G= \frac{\kappa^2}{32\pi c},\quad c = \frac{\kappa^4\mu^2\Lambda}{8(3\lambda-1)^2} .
\end{equation}

In order to study cosmology in this model it is necessary to populate the Universe with matter and radiation. Since we are interested in the phenomenological bounds on such a theory we will introduce a cosmological energy-momentum tensor into the modified Einstein field equations with the simple demand that the standard general relativistic expression is recovered in the infrared limit. Therefore, as it is described in one of our previous papers \cite{Czuchry:2009hz}, we equip our Universe with a hydrodynamic approximation where $p_m$ and $\rho_m$ (pressure and energy density of the  dark plus baryonic matter) are parameters  subject to the continuity equation $\dot{\rho}_m+3H(\rho_m+p_m)=0$. We also include a standard model radiation component through $p_r$ and $\rho_r$ (note that we already have a cosmological constant $\Lambda$ in $S_{db}$), which are also subject to the evolution equation $\dot{\rho}_m+3H(\rho_m+p_m)=0$.

Moreover, we use the \emph{projectability condition} $N=N(t)$ \cite{Horava:2009uw} and the standard FLRW line element: $g_{ij} = a^2(t)\gamma_{ij},\, N_i=0$, where $\gamma_{ij}$ is a
maximally symmetric constant curvature metric:
\begin{equation}\label{eq:friedmann0}
\gamma_{ij}\text{d}x^i\text{d}x^j = \frac{\text{d}r^2}{1-Kr^2}+r^2\text{d}\Omega^2,
\end{equation}
values $K=\{-1,0,1\}$ corresponds to closed, flat, and open Universe, respectively.
On this background
\begin{equation}
K_{ij}=\frac{H}{N}g_{ij}\,,\qquad R_{ij}=\frac{2K}{a^2}g_{ij}\,,\qquad C_{ij}=0\,,
\end{equation}
where $H\equiv \dot a/a$ is the Hubble parameter.

The equations of motion are obtained by varying the action \eqref{eq:action} written in the FLRW metrics with
respect to $N$ and  $a$. After that   lapse is set to one: $N = 1$ (if it was set so at the beginning, standard Friedmann equations would have been obtained) and  terms with  density $\rho$ and pressure $p$  are added). This leads to the Friedmann equations for the projectable Ho\v{r}ava-Lifshitz universe under detailed-balance condition:
\begin{equation}\label{eq:friedmann1}
\left(\frac{\dot{a}}{a}\right)^2 = \frac{\kappa^2}{6(3\lambda-1)}\left[\rho_m+\rho_r\right]+\frac{\kappa^2}{6(3\lambda-1)}\left[\frac{3\kappa^2\mu^2 K^2}{8(3\lambda-1)a^4}+\frac{3\kappa^2\mu^2\Lambda^2}{8(3\lambda-1)}\right]-\frac{\kappa^4\mu^2\Lambda K}{8(3\lambda-1)^2a^2},
\end{equation}
\begin{equation}\label{eq:friedmann2}
\frac{\text{d}}{\text{d}t}\frac{\dot{a}}{a} + \frac{3}{2}\left(\frac{\dot{a}}{a}\right)^2 = -\frac{\kappa^2}{4(3\lambda-1)}[p_m+p_r]-
 \frac{\kappa^2}{4(3\lambda-1)}\left[\frac{\kappa^2\mu^2 K^2}{8(3\lambda-1)a^4}-\frac{3\kappa^2\mu^2\Lambda^2}{8(3\lambda-1)} \right]-\frac{\kappa^4\mu^2\Lambda K}{16(3\lambda-1)^2a^2}.
\end{equation}
We can introduce in the above equations the dark energy parameters, namely energy density $\rho_{DE}$ and pressure density $p_{DE}$, defined as follows:
\begin{align}
\rho_{DE}|_{db}&:=\frac{3\kappa^2\mu^2 K^2}{8(3\lambda-1)a^4}+\frac{3\kappa^2\mu^2\Lambda^2}{8(3\lambda-1)}\label{eq:de},\\
p_{DE}|_{db}&:=\frac{\kappa^2\mu^2 K^2}{8(3\lambda-1)a^4}-\frac{3\kappa^2\mu^2\Lambda^2}{8(3\lambda-1)}.
\end{align}

\subsubsection{Setup for Observational Constraints}
Here we describe the theoretical setup which made it possible to constrain the parameters of the theory. Introducing natural units  $8\pi G  = 1=c$ and taking the IR limit $\lambda=1$ reduces relations \eqref{eq:cosmoconstants} to the following ones:
\begin{equation}\label{eq:cosmoconstants2}
{\kappa^2}=4, \quad \mu^2\Lambda =2.
\end{equation}
Substituting these values to the the Friedmann equation \eqref{eq:friedmann1} leads to:
\begin{equation}\label{eq:friedmann01}
\left(\frac{\dot{a}}{a}\right)^2 = \frac{1}{3}\left(\rho_m+\rho_r\right)+\frac{1}{3}\left(\frac{3 K^2}{2\Lambda a^4}+\frac{3\Lambda}{2}\right)-\frac{ K}{a^2},
\end{equation}
We also define in the IR limit the canonical density parameters for the current universe $a_0=1$ (subscript $0$ indicates the value as measured today)  as follows
\begin{equation}\label{eq:omegadef}
\Omega_m^0 = \frac{\rho_m}{3H_0^2}, \quad \Omega_r^0 = \frac{\rho_r}{3H_0^2}, \quad \Omega_k^0 = -\frac{K}{H_0^2a_0^2},
\end{equation}
where $H_0$ is the Hubble parameter. Using these parameter we rewrite equation \eqref{eq:friedmann01} as follows:
\begin{equation}\label{eq:friedmann001}
\left(\frac{\dot{a}}{a}\right)^2 = H_0^2 \left[\Omega_m^0a^{-3}+\Omega_r^0a^{-4}+\Omega_k^0a^{-2} + \frac{(\Omega_k^0)^2 H_0^2}{2\Lambda}a^{-4} + \frac{\Lambda}{2H_0^2}\right],
\end{equation}
where the last two term correspond to dark energy density parameter obtained from equations (\ref{eq:de}) and (\ref{eq:omegadef}):
\begin{equation}\label{eq:dedb}
\Omega_{DE} = \frac{\rho_{DE}|_{db}}{3H_0^2} = \frac{(\Omega_k^0)^2H_0}{2\Lambda} a^{-4} + \frac{\Lambda}{2H_0^2}.
\end{equation}
In the above Friedmann equation \eqref{eq:friedmann001}
we encounter the term $\Omega_k^2H_0^2/2\Lambda a^{4}$, which is the coefficient of dark radiation in Ho\v{r}ava-Lifshitz. We can conveniently express this in terms of the effective number of neutrino species present in the BBN époque. This is because the dark radiation must have been present during that time and is thus subject to BBN constraints from other experiments. As such, we can obtain a constraint equation for the dark radiation component~\cite{Olive:1999ij, Hagiwara:2002fs}:
\begin{equation}\label{eq:constraintN}
\frac{(\Omega_k^0)^2H_0^2}{2\Lambda} = 0.135\Delta N_\nu\Omega_r^0,
\end{equation}
where $\Delta N_\nu$ is the deviation of the effective number of massless neutrino species from the standard $\Lambda$CDM value ($N_\nu = 3 + \Delta N_\nu$). There are already several bounds on $\Delta N_\nu$ from different experiments. The limits from \cite{Steigman:2005uz,Hagiwara:2002fs}: $-1.7\leq\Delta N_\nu\leq 2.0$ originate from a BBN analysis, and in~\cite{Cyburt:2015mya}, the authors use data from BBN and CMB (Planck) to arrive at an upper limit of $\Delta N_\nu < 0.2$. Moreover, using Planck CMB data, the authors of~\cite{Oldengott:2017tzj} arrive at the limits $-0.32 \leq \Delta N_\nu \leq 0.44$ (TTTEEE+lensing). Other approaches and bounds can be found in ~\cite{Zentner:2001zr, Bean:2002sm}, for example. Through the relation (\ref{eq:constraintN}) we see that $\Delta N_\nu=0$ leads to the possibility of a flat Universe. This is concerning, since it has been established that a flat Ho\v{r}ava-Lifshitz coincide with the standard $\Lambda$CDM model~\cite{Calcagni:2009ar,Kiritsis:2009sh}. Therefore the constraints on the curvature parameter $\Omega_k$ and the BBN neutrino parameter $\Delta N_\nu$ are of utmost importance, and they have been left as free parameters in our analysis. That is, we have not imposed the BBN limit on $\Delta N_\nu$, but the limits we found on this parameter automatically satisfied the BBN constraint.
Now, since all the density parameters have to add up to unity we have:
\begin{equation}\label{eq:db_constr}
\Omega_m^0+\Omega_r^0+\Omega_k^0 + \frac{(\Omega_k^0)^2}{4\cdot 0.135\Delta N_\nu\Omega_r^0} + 0.135\Delta N_\nu\Omega_r^0 = 1,
\end{equation}
and we can rewrite the Friedmann equation (\ref{eq:friedmann001}) as:
\begin{equation}\label{eq:db_fun}
\left(\frac{\dot{a}}{a}\right)^2 = H_0^2 \left[\Omega_m^0a^{-3}+\Omega_r^0a^{-4}+\Omega_k^0a^{-2} + \frac{(\Omega_k^0)^2}{4\cdot 0.135\Delta N_\nu \Omega_r^0} + 0.135\Delta N_\nu\Omega_r^0a^{-4}\right],
\end{equation}
and this is the equation that we have used in our MCMC analysis of detailed balance.

\subsection{Beyond Detailed Balance}
Since there has been a discussion in the literature about whether detailed balance is too restrictive (\cite{Calcagni:2009ar,Kiritsis:2009sh,Wang:2017brl}) it is important to investigate the case when we do not impose this condition. When we relax the detailed balance condition we open up for including more terms into the potential $\mathcal{V}$. In this case, the Friedmann equations can be expressed as~\cite{Charmousis:2009tc,Sotiriou:2010wn}:
\begin{equation}\label{eq:friedmann3}
\left(\frac{\dot{a}}{a}\right)^2 = \frac{2\sigma_0}{3\lambda-1}(\rho_m+\rho_r)+\frac{2}{3\lambda-1}\left[\frac{\Lambda}{2}+\frac{\sigma_3K^2}{6a^4}+\frac{\sigma_4K}{6a^6}\right]+\frac{\sigma_2}{3(3\lambda-1)}\frac{K}{a^2},
\end{equation}
\begin{equation}
\frac{\text{d}}{\text{d}t}\frac{\dot{a}}{a}+\frac{3}{2}\left(\frac{\dot{a}}{a}\right)^2 = -\frac{3\sigma_0}{3\lambda-1}\frac{\rho_r}{3}-\frac{3}{3\lambda-1}\left[-\frac{\Lambda}{2}+\frac{\sigma_3K^2}{18a^4}+\frac{\sigma_4K}{6a^6}\right]+\frac{\sigma_2}{6(3\lambda-1)}\frac{K}{a^2},
\end{equation}
where $\sigma_i$ are arbitrary constants ($\sigma_2<0$) and in analogy with the detailed balance scenario:
\begin{equation}
G = \frac{6\sigma_0}{16\pi}, \quad \frac{\sigma_2}{3(3\lambda-1)} = -1.
\end{equation}
Dark energy and pressure densities may be defined as follows:
\begin{align}
\rho_{DE}|_{bdb}&:=3\left(\frac{\Lambda}{2}+\frac{\sigma_3K^2}{6a^4}+\frac{\sigma_4K}{6a^6}\right)\label{eq:nde},\\
p_{DE}|_{bdb}&:=3\left(-\frac{\Lambda}{2}+\frac{\sigma_3K^2}{6a^4}+\frac{\sigma_4K}{6a^6}\right).
\end{align}

\subsubsection{Setup for Observational Constraints}
Here, we follow the setup for the detailed balance scenario, with $8\pi G = 1=c$, and the same definition of the density parameters ($\Omega_X^0$). Following the notation in \cite{Dutta:2009jn} we introduce similar parameters in the IR limit $\lambda=1$:
\begin{equation}\label{eq:omegas}
\omega_1 = \frac{\Lambda}{2H_0^2}, \quad \omega_3 = \frac{\sigma_3H_0^2\Omega_k^0}{6}, \quad \omega_4 = -\frac{\sigma_4\Omega_k^0}{6},
\end{equation}
and introducing the following Bing Bang nucleosynthesis (BBN) constraint~\cite{Olive:1999ij}:
\begin{equation}\label{eq:bbn}
\omega_3 + \omega_4(1+z_{BBN})^2 = 0.135\Delta N_\nu\Omega_r^0,
\end{equation}
where $z_{BBN}$ is the redshift at BBN ($\sim 4 \times 10^8$). Moreover, we have the following constraint on the density parameters:
\begin{equation}\label{eq:bdb_constr}
\Omega_m^0 + \Omega_r^0 + \Omega_k^0 + \omega_1 + \omega_3 + \omega_4 = 1.
\end{equation}
Using the above two equations to eliminate $\omega_4$ along with the $\sigma$-parameters, we can rewrite the Friedmann equation (\ref{eq:friedmann3}) for the beyond detailed balance scenario as:
\begin{equation}\label{eq:bdb_friedmann}
\left(\frac{\dot{a}}{a}\right)^2 =H_0^2\left[ \Omega_m^0 a^{-3} + \left(\Omega_r^0 + \omega_3\right)a^{-4} + \Omega_k^0 a^{-2} + \omega_1 + \frac{0.135\cdot\Delta N_\nu\Omega_r^0-\omega_3}{(1+z_{BBN})^2}a^{-6}\right].
\end{equation}
This is the equation we have used in our MCMC analysis of the beyond detailed balance scenario. Additionally, the dark energy density parameter can be written as:
\begin{equation}\label{eq:debdb}
\Omega_{DE}^0=\omega_1+\omega_3+\omega_4=\omega_1 + \omega_3 +\frac{0.135\cdot \Delta N_\nu \Omega_r^0 - \omega_3}{(1+z_{BBN})^2}.
\end{equation}

\section{Results and Discussion}
Using the equations (\ref{eq:friedmann1}) and (\ref{eq:friedmann3}) as a starting point, we now want to find the parameter set which best fits the data. To do this we use a large updated data set with CMB (Planck), SN1a, BAO and more (see Appendix for details). In order to do this, we used a Markov-Chain Monte Carlo (MCMC) method, which was evaluated on the CI\'S computer cluster. The parameters are completely unconstrained but are given initial guesses, which speed up computation. We introduced
a Gaussian prior on one parameter: $H_0$, derived from the Hubble constant value in~\cite{Bennett14}, $H_0 = (69.6 \pm 0.7)$~km~s$^{-1}$~Mpc$^{-1}$. That was the only prior imposed on any of the parameters. During every step in the computation, the MCMC method calculates the $\chi^2$, and in the end returns the parameter set which minimised the $\chi^2$ function. This way, we are able to obtain information about the posterior probability distribution without knowing it explicitly. For full details about the data and method, see Appendix.

\subsection{Detailed Balance}
\begin{center}
\begin{figure}[ht]
\includegraphics[scale=0.5]{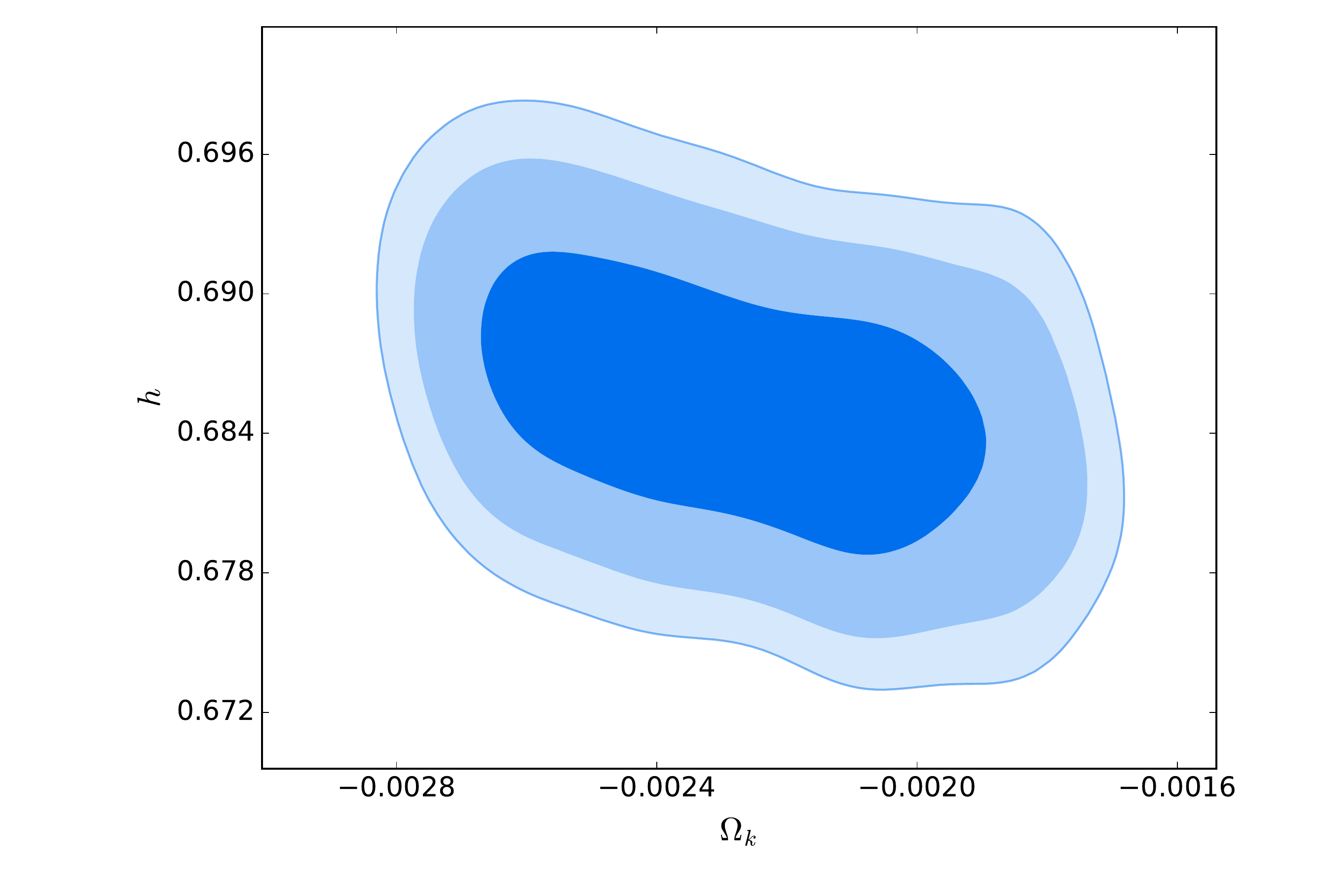}
\caption{$1,2$, and $3\sigma$ contours of the curvature parameter $\Omega_k^0$ and the dimensionless Hubble parameter $h$ under detailed balance. Solid (blue) colour corresponds to the $1\sigma$ limit. Spatial flatness ($\Omega_k^0=0$) is excluded at more than $3\sigma$.}
\label{fig:hOk}
\end{figure}
\end{center}

To derive the constraints on the detailed balance scenario, we used Eq. (\ref{eq:db_fun}) and (\ref{eq:db_fun}) in our MCMC method. As the detailed balance equations are relatively simple there was no need to introduce many constraints, and the parameters $\Delta N_\nu$ has been left completely free in this section. One interesting result is that we find significant evidence for \emph{positive spatial curvature} at $3\sigma$ confidence (see Figure~\ref{fig:hOk}). This is an encouraging prospective observational signal, since any measurement of positive spatial curvature would be a step towards validating this model. Parameters with $1\sigma$ limits are presented in Table~\ref{tab}.

In our analysis, the parameters $H_0$, $\Omega_m^0$, $\Omega_r^0$ all have familiar and reasonable bounds. The parameters which stand out are $\Omega_k^0$, $\Delta N_\nu$ and $\Lambda$. Our initial simulations revealed that the spatial curvature is actually significantly different from zero, and, instead of expressing the parameters $\omega$ in terms of the above parameters as in~\cite{Dutta:2009jn}, we have chosen to leave it as $\omega=\Lambda/2H_0^2$. Therefore there was no need to split the
analysis in $\Omega_k^0 > 0$ and $\Omega_k^0 < 0$. Through Eq.(\ref{eq:constraintN}), a non-zero spatial curvature automatically leads to a non-zero $\Delta N_\nu$. Previous works, like \cite{Dutta:2009jn},  have also used $\Delta N_\nu$ as a free parameter, but with this updated data set and different parametrisation we arrive at different results and parameter bounds (compared, for example, to~\cite{Dutta:2009jn}). In this scenario the value of $\Lambda \sim 0.676\cdot 10^{-35}\textrm{ s}^{-2}\sim 0.836\cdot 10^{-52} \textrm{
    m}^{-2}$ and the cosmological constant in the $\Lambda$CDM model is of the  same order $\Lambda_{\Lambda\textrm{CDM}} \sim 1.11\cdot 10^{-52}$ m$^{-2}$ \cite{planck2015}. There is a marginal difference but the overall magnitude is still much smaller than the lower bound on $\Lambda_\textrm{HL}$ estimated by the energy scales at which Lorentz invariance breaking remains undetected in sub-mm precision tests. Authors of \cite{Blas:2009yd,Pospelov:2010mp,Vernieri:2011aa} provide a rough
estimation for the lower bound of $\Lambda_\textrm{HL}$ to be somewhere in the range from $1\textrm{--}100\ \textrm{meV}^2$ (at least) in natural units, which converts to $10^7\textrm{--}10^{9}\ \textrm{m}^{-2}$  in metric units. The issue of the huge bare value of the cosmological constant demanded by the need for suppression at IR limit of fourth-order operators appearing in the HL action is discussed in the literature, eg. \cite{Vernieri:2011aa,Appignani:2009dy} and references therein,  that is
why the obtained numerical value should be viewed from that perspective. Additionally, through $\mu^2\Lambda = 2$, $\mu$ acquires finite values, unlike the results in~\cite{Dutta:2009jn}, where $\Lambda$ is bound by zero from below, and $\mu$ is unbounded from above.

Moreover, using Eq. (\ref{eq:dedb}) we arrive to the bounds $\Omega_{DE}^0 = 0.686 \pm 0.015$ at $3\sigma$. This is very close to that from Planck 2015~\cite{planck2015} where the tabled value is $\Omega_{DE}^0 = 0.6911 \pm 0.0062$ at $1\sigma$ and to the result from \cite{Frusciante:2015maa}. This is interesting as the authors of the latter use effective field theory and a ``healthy extension'' of HL theory~\cite{Blas:2009qj} to constrain Ho\v{r}ava-Lifshitz cosmology against data. They arrive at the value $\Omega_{DE}^0 = 0.69 \pm 0.02$ at $3\sigma$, which is in close agreement to our findings, although they used not only
background Friedmann equation as we do, but also considered linear perturbations on this background. We also   use a different HL model, so our background equation \eqref{eq:friedmann001} is slightly different from their Friedmann equation (46). Except different
parameters of the theory, \eqref{eq:friedmann001} here contains also curvature term, which arrived to be non-zero, but it does not include massive neutrinos  explicitly, as it is contained in eq. (46) of \eqref{eq:friedmann001}. In contrast, our results differ from those in~\cite{Dutta:2009jn} which uses older data, and this strengthens the case that our results are more accurate, as we use similar background equation.

\begin{table}
\begin{center}
\begin{tabular} {l c c}
 Parameter &  DB: $1\sigma$ limits & BDB: $1\sigma$ limits\\
\hline\\[-3mm]
$\Omega_m^0$ & $0.316\pm 0.0054$ & $0.324\pm 0.0068$\\[1.5mm]
{$\Omega_k^0$} & $(-2.27\pm 0.25)\cdot 10^{-3}$ & $(-3.30\pm 0.20)\cdot 10^{-3}$\\[1.5mm]
$\Omega_r^0$ & $(9.08\pm 0.10)\cdot10^{-4}$ & $\left(9.24^{+0.17}_{-0.21}\right)\cdot 10^{-4}$\\[1.5mm]
$\Omega_{DE}^0$ & $0.686 \pm 0.0053$ & $0.679 \pm 0.0061$\\[1.5mm]
$H_0$ & $68.530 \pm 0.413$ & $69.630 \pm 0.635$\\[1.5mm]
{$\Delta N_\nu$} & $0.155\pm 0.033$ & $0.54^{+0.15}_{-0.21}$\\[1.5mm]
$\Lambda$ & $\left(0.676^{+0.125}_{-0.128}\right)\cdot 10^{-35}$ & $\left(0.691^{+0.147}_{-0.146}\right)\cdot 10^{-35}$\\[1.5mm]
$\sigma_1/H_0^2$ & - & $4.073\pm 0.038$\\[1.5mm]
$\log{\sigma_3H_0^2}$ & - & $0.4^{+1.0}_{-1.7}$\\[1.5mm]
\hline
\end{tabular}
    \caption{Constraints on the parameters from both scenarios: detailed balance (DB) and beyond detailed balance (BDB) (The units of $H_0$ are km$\cdot$s$^{-1}$$\cdot$Mpc$^{-1}$ and of $\Lambda$ are $s^{-2}$). A dash (-) indicates that a parameter is not used in that particular model.}
\label{tab}
\end{center}
\end{table}

\subsection{Beyond Detailed Balance}
\begin{center}
\begin{figure}[ht!]
\centering
\includegraphics[width=0.8\textwidth]{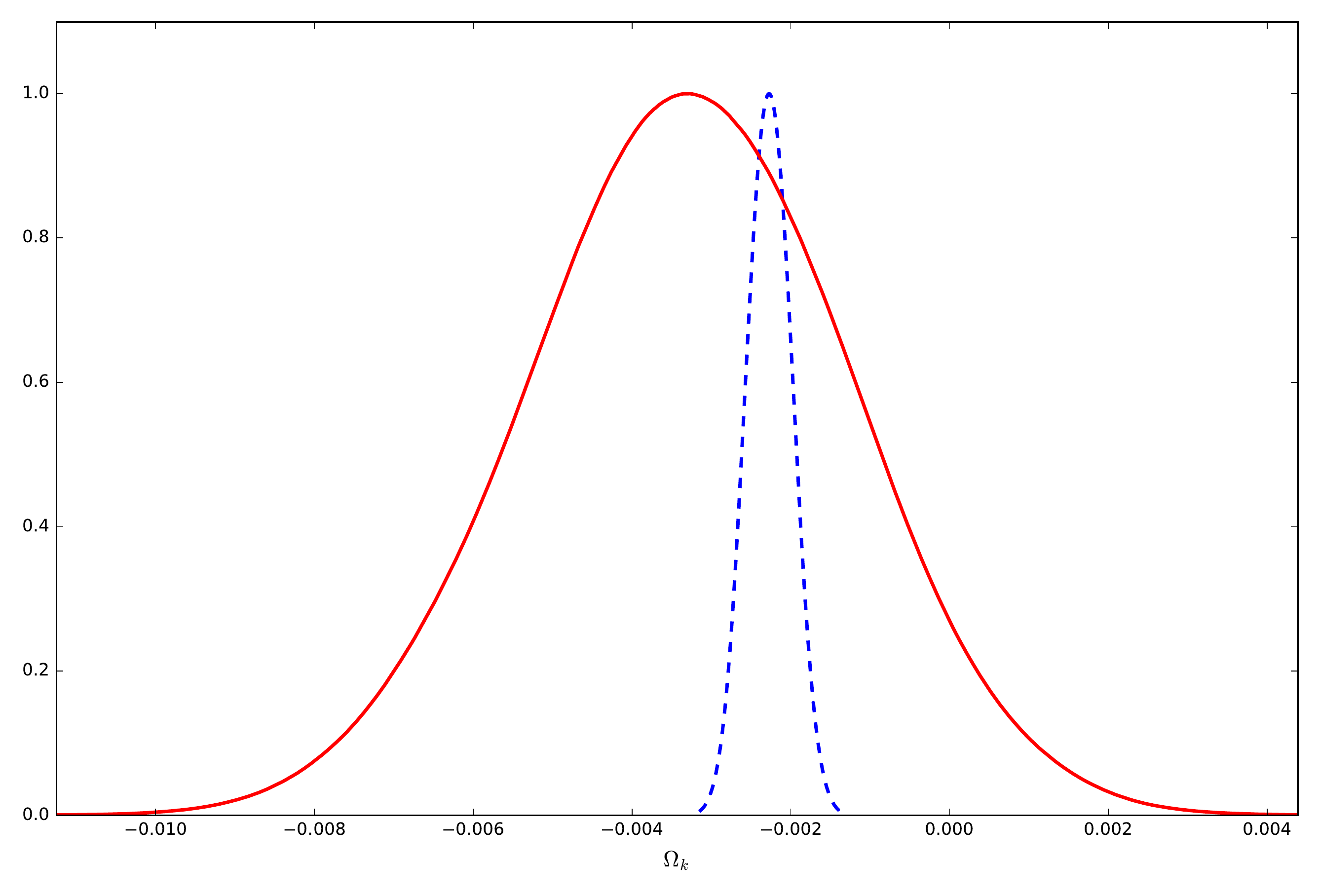}
\caption{Marginalised constraints of the curvature parameter $\Omega_k^0$. The dashed (blue) curve represents the detailed balance scenario while the solid (red) curve shows beyond detailed balance. The detailed balance scenario is completely contained within beyond detailed balance.}
\label{fig:OkOk}
\end{figure}
\end{center}

\begin{center}
\begin{figure}[ht]
\centering
\includegraphics[width=0.8\textwidth]{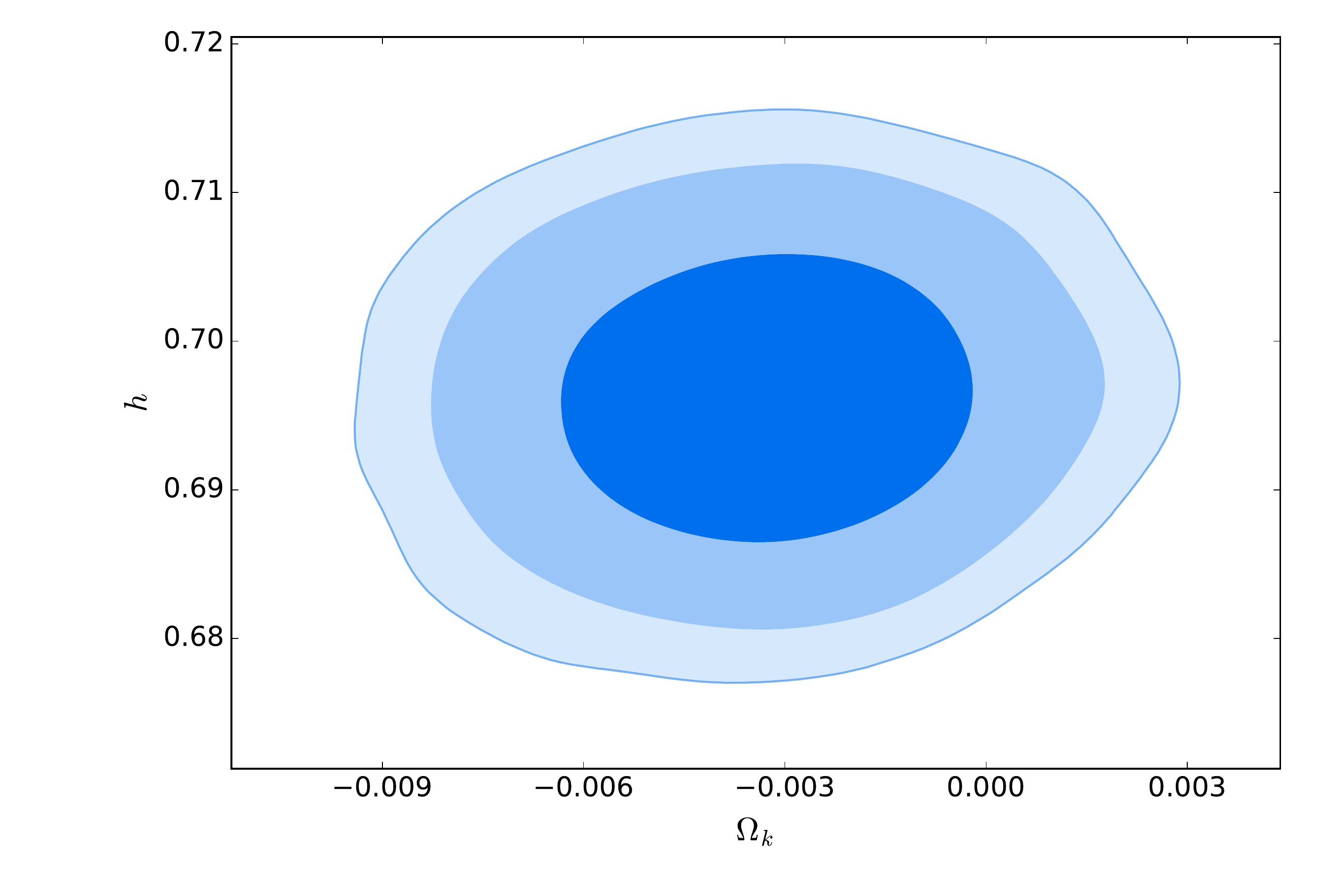}
\caption{$1,2$, and $3\sigma$ contours of the curvature parameter $\Omega_k^0$ and the dimensionless Hubble parameter $h$ in the beyond detailed balance scenario. Solid (blue) colour corresponds to the $1\sigma$ limit. Spatial flatness is excluded at $1\sigma$.}
\label{fig:hOk_beyond}
\end{figure}
\end{center}

To obtain constraints on parameters in the beyond detailed balance scenario we used equations (\ref{eq:bbn}), (\ref{eq:bdb_constr}), and (\ref{eq:bdb_friedmann}) in the same MCMC method as for detailed balance. Here, we introduced another constraint equation in order to make sure that the Hubble parameter is always completely real. From equations (\ref{eq:nde}) and (\ref{eq:bbn})   it is possible to identify a generalised dark energy density. As such, this energy density has to be larger than zero. In this way we obtain real values of the Hubble parameters while only making physical assumptions. The constraint can be written as:
\begin{equation}
\rho_{DE} |_{bdb}= 3H_0^2\left[\omega_1 + \omega_3 a^{-4}+  \omega_4 a^{-6}\right]=3H_0^2\left[\omega_1 + \omega_3 a^{-4}+\frac{0.135\cdot \Delta N_\nu \Omega_r - \omega_3}{(1+z_{BBN})^2}a^{-6}\right] > 0.
\end{equation}
Using the equations and constraints mentioned above we carried out an MCMC analysis of the beyond detailed balance model. All the results are shown in Table~\ref{tab}. Here, it is interesting to note that in this scenario we find evidence for positive spatial curvature at $1\sigma$: $\Omega_k^0 = -0.0033\pm 0.0020$. The contours showing this is plotted against $h$ in Figure~\ref{fig:hOk_beyond}. Here, the marginalised probability for $\Omega_k^0$ in detailed balance is completely contained inside the beyond detailed balance scenario, away from the best-fit value. This can be seen in Figure~\ref{fig:OkOk}.

\begin{center}
\begin{figure}[ht]
\centering
\includegraphics[width=0.8\textwidth]{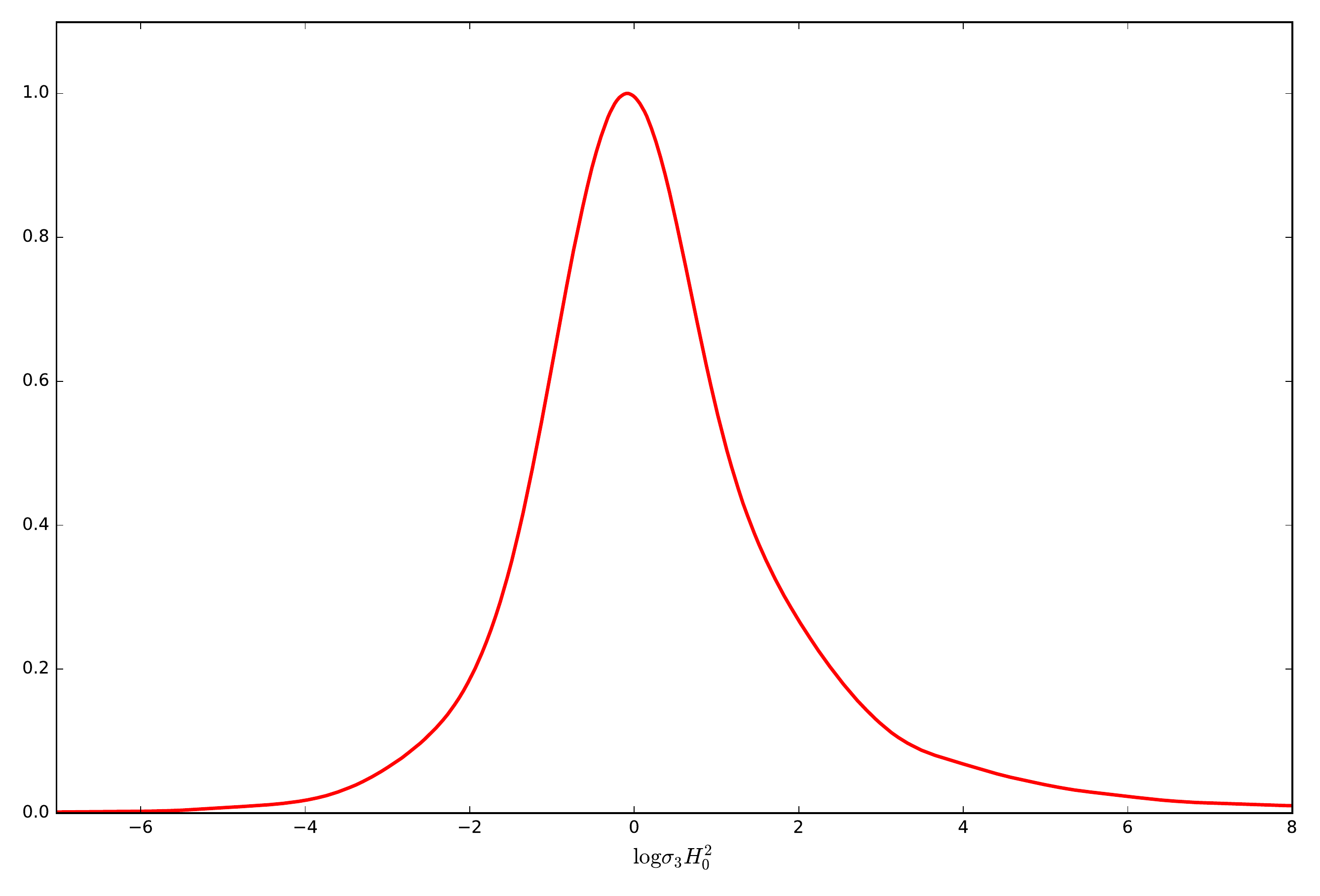}
\caption{Marginalised constraints of the parameter $\log{(\sigma_3H_0^2)}$. The slightly elongated tail on the right hand side shifts the mean away from $0$.}
\label{fig:logsigma3norm}
\end{figure}
\end{center}

Moreover, the parameter $\sigma_3H_0^2$ seems to acquire a log-normal distribution, and the two-sided confidence intervals for this parameter must be considered unreliable. Because of the log-normal distribution, however, it was possible to find limits on $\log{\sigma_3H_0^2}$, which is shown in Figure~\ref{fig:logsigma3norm}. The errors on this parameter are very large, which is an artefact of the log-normal distribution which has long tails on both ends.  We express the two $\sigma$ parameters as $\sigma_1/H_0^2$ and $\sigma_3H_0^2$ (see Eq. (\ref{eq:omegas})) as there is no need to specify units or value for $H_0$.

\begin{center}
\begin{figure}[ht]
\centering
\includegraphics[width=0.8\textwidth]{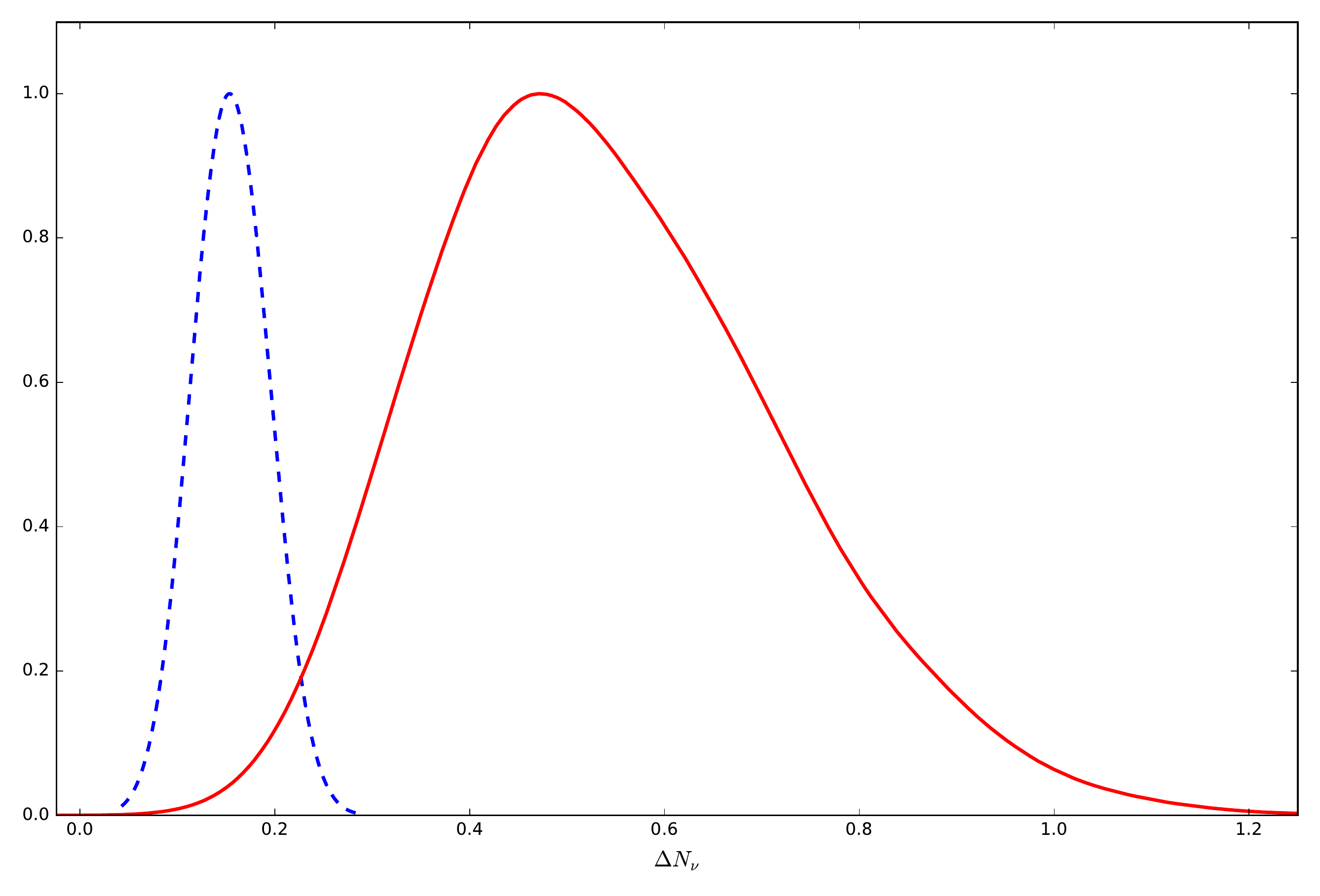}
\caption{Marginalised constraint of the parameter $\Delta N_\nu$. The dashed (blue) curve represents the detailed balance scenario, and the solid (red) shows beyond detailed balance. The two scenarios only shows a small amount of overlap.}
\label{fig:DeltaNDeltaN}
\end{figure}
\end{center}

It is also worth noting that it is possible to differentiate between the two scenarios by looking at the neutrino species parameter $\Delta N_\nu$. As can be seen in Figure~\ref{fig:DeltaNDeltaN}, only some of the marginalised probability of the detailed balance model is contained within the range of beyond detailed balance, and a large part of the curve lies outside. Thus, $\Delta N_\nu$ is a promising parameter to help differentiate between the two scenarios. Remarkably, the lower bound of $\Delta N_\nu$ from Table~\ref{tab} is in conflict with the bound ($\Delta N_\nu < 0.2$) derived from BBN and Planck CMB (mentioned in Section 2.2.1). In fact, looking at the $2\sigma$ and $3\sigma$ bounds on $\Delta N_\nu$ ($0.21, 0.17$) we see that our analysis yields a significantly different result at more than $2\sigma$ confidence.
Using Eq. (\ref{eq:debdb}) we find the bounds $\Omega_{DE}^0 = 0.678^{+ 0.017}_{- 0.019}$ at $3\sigma$ confidence level, which is close to the value from Planck 2015 ($\Omega_{DE}^0 = 0.6911 \pm 0.0062$ at $1\sigma$), but not as close as for the detailed balance scenario. This is mainly due to different estimated values of $\Lambda/2H_0^2$, makes up the main part of $\Omega_{DE}^0$ in both scenarios. Beyond detailed balance parameters at 3$\sigma$ are $\omega_1 = 0.679^{+0.017}_{-0.019}$ (the leading term in $\Omega_{DE}^0$),
$\omega_3 = (2.05^{+4.38}_{-2.04})\cdot 10^{-6}$ and
$\omega_4 (1+z_{BBN})^2 = (4.18\cdot 10^{-6})^{+0.0056}_{-0.0064}$ (the errors on $\omega_3$ and $\omega_4$ are very large, so only order is relevant here).

\section{Conclusions}
Using an updated cosmological data set we calculated new observational constraints on cosmological parameters of different formulations of Ho\v{r}ava-Lifshitz gravity. Unlike in the previous approaches \cite{Dutta:2009jn} we use a different parametrisation of our parameters, the most important one is treating $\Omega_k^0$ (together with $\Delta N_\nu$) as completely free parameter.  This allowed  us to avoid the splitting of considerations and calculations into two separate cases of negative and positive curvature parameters that was used before.

We investigated two basic Ho\v{r}ava-Lifshitz scenarios, the one with detailed balance condition imposed in the action and the other one, the so called Sotiriou-Visser-Weinfurtner (SVW) generalization~\cite{Sotiriou:2009bx}, which relaxes  this condition and assumes higher order, i.e. cubic terms in the action. We found very remarkable results, namely detailed balance scenario exhibits positive spatial curvature to more than  $3\sigma$ whereas for the SVW  generalization there is  evidence for positive spatial curvature at 1$\sigma$, which could be a smoking gun  for observations. Some of our parameters differ quite a lot from previous works, but as we have some different assumptions (for example, in~\cite{Dutta:2009jn} the parameter $\omega_3$ is assumed to be positive for convenience), different values are to be expected.

The best fit value of the parameter $\Delta N_\nu$ (kept as a completely free parameter of the theory) at 1$\sigma$ is definitely positive and within BBN limits, as well as within some of the bounds from CMB. Since there is a variety of papers with different bounds on $\Delta N_\nu$ available, it is unclear which one should compare to. In general, more data is needed to make a final verdict on this issue, and we aim to address this in future work as it becomes available. Our analysis, however, excludes from both scenarios the zero-curvature flat universe and corresponds to the  non-zero positive cosmological constant $\Lambda$.

Most values of the parameters of the theory are similar as in  $\Lambda$CDM model. For example, our values for $\Omega_m = 0.316 \pm 0.0054 (0.324 \pm 0.0068)$ for detailed balance (beyond detailed balance) are both within $1\sigma$ of $\Lambda$CDM~\cite{planck2015}. We are also close to the dark energy density parameter, as discussed in Section 3.2. However, what sets the HL model apart is the non-zero curvature parameter $\Omega_k$. We found it to be different from zero to $1\sigma$
($3\sigma$). According to Planck 2015 (when including BAO's) this parameter is a Gaussian centered around zero~\cite{planck2015}. As our analysis also includes BAO's, further investigation of this parameter may be the way to finally exclude one of these models. Naively, one must still prefer the $\Lambda$CDM model, as it has fewer parameters than HL and fits the data well. However, bearing in mind the theoretical basis of HL cosmology (candidate for a UV-complete theory of gravity) there
are, in our opinion, plenty of reasons to keep investigating this model.

It is also worth mentioning that different experiments measuring the Hubble constant $H_0$ provide different values (the so-called $H_0$ tension). Our results: $68.5 \,\pm \,0.4$ and $69.6 \,\pm \,0.6$ km s$^{-1}$ Mpc$^{-1}$ for detailed balance and beyond detailed balance, respectively, lie close to the Planck 2018 result $67.27 \,\pm \,0.6 $ km s$^{-1}$ Mpc$^{-1}$ in mean, but our $1\sigma$ confidence levels overlap with those from e.g. strong gravitational lensing,
$71.9^{+2.4}_{-3}$~\cite{Bonvin} and lies close to those of cosmic ladder experiments like the Sh0es project, where the Hubble constant was found to be $73.8 \,\pm \,2.4$ km s$^{-1}$ Mpc$^{-1}$~\cite{Riess}. However, this experiment used a Gaussian prior on $H_0$ which we have also done, so this agreement might be an artifact of this choice. In short, $H_0$ is not a good parameter for differentiating between HL models or to cast any light on the $H_0$ tension.

In this paper we were mostly interested in the IR regime of the HL theory and wanted to know if the values of cosmological observables predicted within the considered models fit  observational data, and also whether they may be used to distinguish HL cosmology from GR, even in the low-energy limit.  Therefore we set the value of the running coupling constant $\lambda$ to its IR value: $\lambda=1$. There are works \cite{Dutta:2010jh} where authors estimate the value of $\lambda$ using
similar low-energy data sources (however, our data is more recent), providing that this value  is restricted to $|\lambda-1|\leq 0.02$ at $1\sigma$.  At the moment we are performing  calculations within more general formulation of the theory \cite{Blas:2009qj} and wider observational data, with more high-energy sources, where we also investigate the impact of  running of $\lambda$ on other results.

\vspace{5mm}
{\bf Acknowledgments}
The authors wish to thank Mariusz P. D\c{a}browski (University of Szczecin, NCBJ) for helpful and inspiring discussions. N.A.N.  wishes to thank Vincenzo Salzano (University of Szczecin) for valuable discussion and assistance with the MCMC method. N.A.N.  was supported under the Polish National Science Centre Grant DEC-2012/06/A/ST2/00395. The use of the CI\'S supercomputer at the National Centre for Nuclear Research is greatly appreciated.

\section*{Appendix: The Method}\label{data}
In order to estimate the values of the parameters present in Ho\v{r}ava-Lifshitz we used a large updated cosmological data set. The data used includes expansion rates of elliptical and lenticular galaxies~\cite{Moresco15}, Type Ia Supernovae~\cite{JLA}, Baryon Acoustic Oscillations~\cite{WiggleZ,SDSS12,Lyman}, Cosmic Microwave Background~\cite{planck2015} and priors on the Hubble parameter~\cite{Bennett14}. We used the parallelised Markov-Chain Monte Carlo (MCMC) code developed in Mathematica and used in many papers, for example~\cite{Nilsson:2016rsv,Salzano:2016pny}. There are many cosmological MCMC codes readily available, for example CosmoMC~\cite{Lewis:2002ah} and Monte Python~\cite{2013JCAP...02..001A}, but the advantages of our code of choice is that it is easy to add new data, and it is also simple to modify. Things such as the cosmological model, statistical method, and parameters used can easily be changed. Moreover, thanks to the flexible way in which the code is written, introducing more exotic models with varying $c$ or $G$ is also straightforward. Even though Mathematica code is generally slower than C or FORTRAN, for example, the simplicity and transparency of the code makes up for this small drawback.

All expressions below are written in the flat case ($\Omega_k^0 = 0$) for simplicity. However, $\Omega_k^0$ is left as a free parameter in the MCMC analysis, and thus all equations extend to the more general case of~\cite{Hogg}.

With this in mind, the expression for the comoving distance is:
\begin{equation}\label{eq:DM_k}
D_{M}(z) = \begin{cases} \frac{D_{H}}{\sqrt{\Omega_{k}}} \sinh \left( \sqrt{\Omega_{k}} \frac{D_{C}(z)}{D_{H}} \right) &\mbox{for } \Omega_{k}^0 > 0\\
D_{C}(z) &\mbox{for } \Omega_{k}^0 = 0 \\
\frac{D_{H}}{\sqrt{|\Omega_{k}^0|}} \sin \left( \sqrt{|\Omega_{k}^0|} \frac{D_{C}(z)}{D_{H}} \right) &\mbox{for } \Omega_{k}^0 < 0 \, ,
\end{cases}
\end{equation}
where $D_{H} = c_{0} / H_{0}$ is the Hubble distance, $D_{C}(z) = D_{H} \int^{z}_{0} dz' / \mathcal{E}(z')$ is the line-of-sight comoving distance, and $\mathcal{E}(z) = H(z)/H_{0}$.  Also, luminosity distance ($D_L(z)$) and angular diameter distance ($D_A(z)$) are given by:
\begin{eqnarray}
D_{L}(z) &=& (1+z) D_{M}(z)\; , \\
\label{angdist1}D_{A}(z) &=& \frac{D_{M}(z)}{1+z} \; .
\end{eqnarray}

\subsection*{Hubble data}
For Hubble parameter data, we use the compilation from~\cite{Moresco15}, which is derived from the evolution of elliptical and lenticular galaxies at redshifts $0<z<1.97$. The expression for $\chi^2_H$ is:
\begin{equation}\label{eq:hubble_data}
\chi^2_{H}= \sum_{i=1}^{24} \frac{\left( H(z_{i},\boldsymbol{\theta})-H_{obs}(z_{i}) \right)^{2}}{\sigma^2_{H}(z_{i})} \; ,
\end{equation}
where $\boldsymbol{\theta}$ is a vector containing the parameters of the model, $H_{obs}(z_{i})$ are the measured values of the Hubble constant and $\sigma_{H}(z_{i})$ are the corresponding observational errors. We will also add a prior on the Hubble constant from~\cite{Bennett14}, $H_0=69.6 \pm 0.7$ km s$^{-1}$ Mpc$^{-1}$.

\subsection*{Type Ia Supernovae}
We made use of the updated JLA compilation (Joint Light-Curve Analysis) data for Type Ia supernovae (SneIa)~\cite{JLA} at redshifts $0<z<1.39$. In this case, the $\chi_{SN}^2$ is:
\begin{equation}
\chi^2_{SN} = \Delta \boldsymbol{\mu} \; \cdot \; \mathbf{C}^{-1}_{SN} \; \cdot \; \Delta  \boldsymbol{\mu} \; ,
\end{equation}
where $\Delta\boldsymbol{\mu} = \mu_{theo} - \mu_{obs}$ is the difference between theoretical and observational values of the distance modulus $\mu$. $\mathbf{C}_{SN}$ is the total covariance matrix.
The distance modulus is written as:
\begin{equation}\label{eq:m_jla}
\mu(z,\boldsymbol{\theta}) = 5 \log_{10} [ D_{L}(z, \boldsymbol{\theta}) ] - \alpha X_{1} + \beta \mathcal{C} + \mathcal{M}_{B} \; .
\end{equation}
Here, $X_1$ parametrises the shape of the supernova light-curve, $\mathcal{C}$ is the colour, and $\mathcal{M}_B$ is a nuisance parameter~\cite{JLA}, which together with the weighting parameters $\alpha$ and $\beta$ are included in $\boldsymbol{\theta}$. $D_{L}$ is the luminosity distance which we write as:
\begin{equation}\label{eq:dL}
D_{L}(z, \boldsymbol{\theta})  = \frac{1+z}{H_{0}}\int_{0}^{z} \frac{\mathrm{d}z'}{\mathcal{E}(z',\boldsymbol{\theta})} \; .
\end{equation}
Here, and only in this Supernova analysis, we do specify $H_0 = 70$ km/s Mpc$^{-1}$~\cite{JLA}.

\subsection*{Baryon Acoustic Oscillations}
The total $\chi^2$ for Baryon Acoustic Oscillations is given by:
\begin{equation}
\chi^2_{BAO} = \Delta \boldsymbol{\mathcal{F}}^{BAO} \; \cdot \; \mathbf{C}^{-1}_{BAO} \; \cdot \; \Delta  \boldsymbol{\mathcal{F}}^{BAO} \; ,
\end{equation}
where $\boldsymbol{\mathcal{F}}^{BAO}$ is different from survey to survey. In this work, we used the WiggleZ Dark Energy Survey with redshifts $z=\{0.44,0.6,0.73\}$~\cite{WiggleZ}.
For this analysis the acoustic parameter and the Alcock-Paczynski distortion parameter are of interest. The acoustic parameter is defined as follows:
\begin{equation}\label{eq:AWiggle}
A(z, \boldsymbol{\theta}) = 100  \sqrt{\Omega_{m}^0 \, h^2} \frac{D_{V}(z,\boldsymbol{\theta})}{z} \, ,
\end{equation}
and the Alcock-Paczynski parameter is:
\begin{equation}\label{eq:FWiggle}
F(z, \boldsymbol{\theta}) = (1+z)  D_{A}(z,\boldsymbol{\theta})\, H(z,\boldsymbol{\theta}) \, ,
\end{equation}
where $D_{A}$ is the angular diameter distance, which is Eq.~(\ref{angdist1}) in the case of $\Omega_k^0 = 0$:
\begin{equation}\label{eq:dA}
D_{A}(z, \boldsymbol{\theta} )  = \frac{1}{H_{0}} \frac{1}{1+z} \ \int_{0}^{z} \frac{\mathrm{d}z'}{\mathcal{E}(z',\boldsymbol{\theta})} \; ,
\end{equation}
and $D_{V}$ is the geometric mean of the physical angular diameter distance $D_A$ and the Hubble function $H(z)$. It reads as:
\begin{equation}\label{eq:dV}
D_{V}(z, \boldsymbol{\theta} )  = \left[ (1+z)^2 D^{2}_{A}(z,\boldsymbol{\theta}) \frac{z}{H(z,\boldsymbol{\theta})}\right]^{1/3}.
\end{equation}

Included in the Baryon Acoustic Oscillation analysis is also data from Sloan Digital Sky Survey (SDSS-III) Baryon Oscillation Spectroscopic Survey (BOSS) DR$12$~\cite{SDSS12}. It can be written as:
\begin{equation}
D_{M}(z) \frac{r^{mod}_{s}(z_{d})}{r_{s}(z_{d})} \qquad \mathrm{and} \qquad H(z) \frac{r_{s}(z_{d})}{r^{mod}_{s}(z_{d})} \,
\end{equation}
Here, $r_s(z_d)$ represents the sound horizon at the \emph{dragging redshift} $z_d$. $r^{mod}_s(z_d)$ is the same horizon, but evaluated for the given cosmological model. Here, it is used that $r^{mod}_s(z_d) = 147.78$ Mpc as in~\cite{SDSS12}. A good approximation of the sound horizon can be found in~\cite{Eisenstein}:
\begin{equation}\label{eq:zdrag}
z_{d} = \frac{1291 (\Omega_{m}^0 \, h^2)^{0.251}}{1+0.659(\Omega_{m}^0 \, h^2)^{0.828}} \left[ 1+ b_{1} (\Omega_{b}^0 \, h^2)^{b2}\right]\; ,
\end{equation}
where
\begin{eqnarray}
b_{1} &=& 0.313 (\Omega_{m}^0 \, h^2)^{-0.419} \left[ 1+0.607 (\Omega_{m}^0 \, h^2)^{0.6748}\right], \nonumber \\
b_{2} &=& 0.238 (\Omega_{m}^0 \, h^2)^{0.223}.
\end{eqnarray}

The sound horizon $r_s$ can then be defined as:
\begin{equation}\label{eq:soundhor}
r_{s}(z, \boldsymbol{\theta}) = \int^{\infty}_{z} \frac{c_{s}(z')}{H(z',\boldsymbol{\theta})} \mathrm{d}z'\, ,
\end{equation}
and the sound speed is given by:
\begin{equation}\label{eq:soundspeed}
c_{s}(z) = \frac{1}{\sqrt{3(1+\overline{R}_{b}\, (1+z)^{-1})}} \; ,
\end{equation}
and
\begin{equation}
\overline{R}_{b} = 31500 \Omega_{b}^0 \, h^{2} \left( T_{CMB}/ 2.7 \right)^{-4}\; ,
\end{equation}
with $T_{CMB} = 2.726$ K.

Ending the Baryon Acoustic Oscillation analysis, considered data from the Quasar-Lyman $\alpha$ Forest from Sloan Digital Sky Survey - Baryon Oscillation Spectroscopic Survey DR$11$~\cite{Lyman}:
\begin{eqnarray}
\frac{D_{A}(z=2.36)}{r_{s}(z_{d})} &=& 10.8 \pm 0.4\; , \\
\frac{1}{H(z=2.36) r_{s}(z_{d})}  &=& 9.0 \pm 0.3\; .
\end{eqnarray}

With the contributions mentioned throughout this section, the total $\chi^2$ for Baryon Acoustic Oscillations will be $\chi^{2}_{BAO} = \chi^{2}_{WiggleZ} + \chi^{2}_{BOSS} + \chi^{2}_{Lyman}$.

\subsection*{Cosmic Microwave Background}
In this analysis, we write the $\chi^2$ for the Cosmic Microwave Background (CMB) in the following way:
\begin{equation}
\chi^2_{CMB} = \Delta \boldsymbol{\mathcal{F}}^{CMB} \; \cdot \; \mathbf{C}^{-1}_{CMB} \; \cdot \; \Delta  \boldsymbol{\mathcal{F}}^{CMB} \; .
\end{equation}
Here, $\boldsymbol{\mathcal{F}}^{CMB}$ is a vector quantity given in~\cite{WangWang}, which summarises the information available in the full power spectrum of the Cosmic Microwave Background from the 2015 Planck data release~\cite{planck2015}. $\boldsymbol{\mathcal{F}}^{CMB}$ contains the Cosmic Microwave Background shift parameters and the baryonic density parameter.
The shift parameters are:
\begin{eqnarray}
R(\boldsymbol{\theta}) &\equiv& \sqrt{\Omega_m^0 H^2_{0}} r(z_{\ast},\boldsymbol{\theta}) \nonumber \\
l_{a}(\boldsymbol{\theta}) &\equiv& \pi \frac{r(z_{\ast},\boldsymbol{\theta})}{r_{s}(z_{\ast},\boldsymbol{\theta})}\, ,
\end{eqnarray}
whereas the baryonic density parameter is simply $\Omega_b^0 \, h^{2}$. As mentioned before, $r_{s}$ is the comoving sound horizon at the photon-decoupling redshift $z_{\ast}$, which is given by~\cite{Hu}:
\begin{equation}{\label{eq:zdecoupl}}
z_{\ast} = 1048 \left[ 1 + 0.00124 (\Omega_{b}^0 h^{2})^{-0.738}\right] \left(1+g_{1} (\Omega_{m}^0 h^{2})^{g_{2}} \right) \, ,
\end{equation}
with:
\begin{eqnarray}
g_{1} &=& \frac{0.0783 (\Omega_{b}^0 h^{2})^{-0.238}}{1+39.5(\Omega_{b}^0 h^{2})^{-0.763}}\; , \\
g_{2} &=& \frac{0.560}{1+21.1(\Omega_{b}^0 h^{2})^{1.81}} \, ;
\end{eqnarray}
and $r$ is the comoving distance:
\begin{equation}
r(z, \boldsymbol{\theta_{b}} )  = \frac{1}{H_{0}} \int_{0}^{z} \frac{\mathrm{d}z'}{\mathcal{E}(z',\boldsymbol{\theta})} \mathrm{d}z'\; .
\end{equation}

All the methods mentioned above all contribute to the total $\chi^2$, and the function to minimise is: $\chi^2_{tot} = \chi^{2}_{H_{0}} + \chi^{2}_{H} + \chi^{2}_{SN} + \chi^{2}_{WiggleZ} + \chi^{2}_{BOSS} + \chi^{2}_{Lyman} + \chi^{2}_{CMB}$.

\bibliographystyle{unsrtnat}
\bibliography{mybibfile}

\begin{thebibliography}{77}
\providecommand{\natexlab}[1]{#1}
\providecommand{\url}[1]{\texttt{#1}}
\expandafter\ifx\csname urlstyle\endcsname\relax
  \providecommand{\doi}[1]{doi: #1}\else
  \providecommand{\doi}{doi: \begingroup \urlstyle{rm}\Url}\fi

\bibitem[Quevedo(2016)]{Quevedo:2016tbh}
Fernando Quevedo.
\newblock {Is String Phenomenology an Oxymoron?}
\newblock \emph{[arXiv:1612.01569]}, 2016.

\bibitem[Girelli et~al.(2012)Girelli, Hinterleitner, and Major]{Girelli:2012ju}
Florian Girelli, Franz Hinterleitner, and Seth Major.
\newblock {Loop Quantum Gravity Phenomenology: Linking Loops to Observational
  Physics}.
\newblock \emph{SIGMA}, 8:\penalty0 098, 2012.
\newblock \doi{10.3842/SIGMA.2012.098}.

\bibitem[Kifune(1999)]{Kifune:1999ex}
Tadashi Kifune.
\newblock {Invariance violation extends the cosmic ray horizon?}
\newblock \emph{Astrophys. J.}, 518:\penalty0 L21--L24, 1999.
\newblock \doi{10.1086/312057}.

\bibitem[Protheroe and Meyer(2000)]{Protheroe:2000hp}
R.~J. Protheroe and H.~Meyer.
\newblock {An Infrared background TeV gamma-ray crisis?}
\newblock \emph{Phys. Lett.}, B493:\penalty0 1--6, 2000.
\newblock \doi{10.1016/S0370-2693(00)01113-8}.

\bibitem[Fairbairn et~al.(2014)Fairbairn, Nilsson, Ellis, Hinton, and
  White]{Fairbairn:2014kda}
Malcolm Fairbairn, Albin Nilsson, John Ellis, Jim Hinton, and Richard White.
\newblock {The CTA Sensitivity to Lorentz-Violating Effects on the Gamma-Ray
  Horizon}.
\newblock \emph{JCAP}, 1406:\penalty0 005, 2014.
\newblock \doi{10.1088/1475-7516/2014/06/005}.

\bibitem[Abbott et~al.(2016)]{Abbott:2016blz}
Benjamin~P. Abbott et~al.
\newblock {Observation of Gravitational Waves from a Binary Black Hole Merger}.
\newblock \emph{Phys. Rev. Lett.}, 116\penalty0 (6):\penalty0 061102, 2016.
\newblock \doi{10.1103/PhysRevLett.116.061102}.

\bibitem[Abbott et~al.(2017{\natexlab{a}})]{TheLIGOScientific:2017qsa}
Benjamin~P. Abbott et~al.
\newblock {GW170817: Observation of Gravitational Waves from a Binary Neutron
  Star Inspiral}.
\newblock \emph{Phys. Rev. Lett.}, 119\penalty0 (16):\penalty0 161101,
  2017{\natexlab{a}}.
\newblock \doi{10.1103/PhysRevLett.119.161101}.

\bibitem[Troja et~al.(2017)]{Troja:2017nqp}
E.~Troja et~al.
\newblock {The X-ray counterpart to the gravitational wave event GW 170817}.
\newblock \emph{Nature}, 551:\penalty0 71--74, 2017.
\newblock \doi{10.1038/nature24290}.
\newblock [Nature551,71(2017)].

\bibitem[Abbott et~al.(2017{\natexlab{b}})]{Monitor:2017mdv}
Benjamin~P. Abbott et~al.
\newblock {Gravitational Waves and Gamma-rays from a Binary Neutron Star
  Merger: GW170817 and GRB 170817A}.
\newblock \emph{Astrophys. J.}, 848\penalty0 (2):\penalty0 L13,
  2017{\natexlab{b}}.
\newblock \doi{10.3847/2041-8213/aa920c}.

\bibitem[Brans and Dicke(1961)]{PhysRev.124.925}
C.~Brans and R.~H. Dicke.
\newblock Mach's principle and a relativistic theory of gravitation.
\newblock \emph{Phys. Rev.}, 124:\penalty0 925--935, Nov 1961.
\newblock \doi{10.1103/PhysRev.124.925}.

\bibitem[Blas and Sanctuary(2011)]{Blas:2011zd}
Diego Blas and Hillary Sanctuary.
\newblock {Gravitational Radiation in Hořava Gravity}.
\newblock \emph{Phys. Rev.}, D84:\penalty0 064004, 2011.
\newblock \doi{10.1103/PhysRevD.84.064004}.

\bibitem[Emir~Gümrükçüoğlu et~al.(2018)Emir~Gümrükçüoğlu, Saravani,
  and Sotiriou]{Gumrukcuoglu:2017ijh}
A.~Emir~Gümrükçüoğlu, Mehdi Saravani, and Thomas~P. Sotiriou.
\newblock {Hořava gravity after GW170817}.
\newblock \emph{Phys. Rev.}, D97\penalty0 (2):\penalty0 024032, 2018.
\newblock \doi{10.1103/PhysRevD.97.024032}.

\bibitem[Oost et~al.(2018)Oost, Mukohyama, and Wang]{Oost:2018tcv}
Jacob Oost, Shinji Mukohyama, and Anzhong Wang.
\newblock {Constraints on Einstein-aether theory after GW170817}.
\newblock 2018.

\bibitem[Abbott et~al.(2018{\natexlab{a}})]{Abbott:2017tlp}
Benjamin~P. Abbott et~al.
\newblock {First search for nontensorial gravitational waves from known
  pulsars}.
\newblock \emph{Phys. Rev. Lett.}, 120\penalty0 (3):\penalty0 031104,
  2018{\natexlab{a}}.
\newblock \doi{10.1103/PhysRevLett.120.031104}.

\bibitem[Abbott et~al.(2018{\natexlab{b}})]{Abbott:2018utx}
Benjamin~P. Abbott et~al.
\newblock {A Search for Tensor, Vector, and Scalar Polarizations in the
  Stochastic Gravitational-Wave Background}.
\newblock 2018{\natexlab{b}}.

\bibitem[Arkani-Hamed et~al.(1998)Arkani-Hamed, Dimopoulos, and
  Dvali]{ArkaniHamed:1998rs}
Nima Arkani-Hamed, Savas Dimopoulos, and G.~R. Dvali.
\newblock {The Hierarchy problem and new dimensions at a millimeter}.
\newblock \emph{Phys. Lett.}, B429:\penalty0 263--272, 1998.
\newblock \doi{10.1016/S0370-2693(98)00466-3}.

\bibitem[Dvali(2010)]{Dvali:2007hz}
Gia Dvali.
\newblock {Black Holes and Large N Species Solution to the Hierarchy Problem}.
\newblock \emph{Fortsch. Phys.}, 58:\penalty0 528--536, 2010.
\newblock \doi{10.1002/prop.201000009}.

\bibitem[Hořava(2009)]{Horava:2009uw}
Petr Hořava.
\newblock {Quantum Gravity at a Lifshitz Point}.
\newblock \emph{Phys. Rev.}, D79:\penalty0 084008, 2009.
\newblock \doi{10.1103/PhysRevD.79.084008}.

\bibitem[Hořava and Melby-Thompson(2010)]{Horava:2010zj}
Petr Hořava and Charles~M. Melby-Thompson.
\newblock {General Covariance in Quantum Gravity at a Lifshitz Point}.
\newblock \emph{Phys. Rev.}, D82:\penalty0 064027, 2010.
\newblock \doi{10.1103/PhysRevD.82.064027}.

\bibitem[Mukohyama(2010)]{Mukohyama:2010xz}
Shinji Mukohyama.
\newblock {Hořava-Lifshitz Cosmology: A Review}.
\newblock \emph{Class. Quant. Grav.}, 27:\penalty0 223101, 2010.
\newblock \doi{10.1088/0264-9381/27/22/223101}.

\bibitem[Kiritsis and Kofinas(2009)]{Kiritsis:2009sh}
Elias Kiritsis and Georgios Kofinas.
\newblock {Hořava-Lifshitz Cosmology}.
\newblock \emph{Nucl. Phys.}, B821:\penalty0 467--480, 2009.
\newblock \doi{10.1016/j.nuclphysb.2009.05.005}.

\bibitem[Calcagni(2009)]{Calcagni:2009ar}
Gianluca Calcagni.
\newblock {Cosmology of the Lifshitz universe}.
\newblock \emph{JHEP}, 09:\penalty0 112, 2009.
\newblock \doi{10.1088/1126-6708/2009/09/112}.

\bibitem[Saridakis(2010)]{Saridakis:2009bv}
Emmanuel~N. Saridakis.
\newblock {Hořava-Lifshitz Dark Energy}.
\newblock \emph{Eur. Phys. J.}, C67:\penalty0 229--235, 2010.
\newblock \doi{10.1140/epjc/s10052-010-1294-6}.

\bibitem[Brandenberger(2009)]{Brandenberger:2009yt}
Robert Brandenberger.
\newblock {Matter Bounce in Hořava-Lifshitz Cosmology}.
\newblock \emph{Phys. Rev.}, D80:\penalty0 043516, 2009.
\newblock \doi{10.1103/PhysRevD.80.043516}.

\bibitem[Czuchry(2011{\natexlab{a}})]{Czuchry:2010vx}
Ewa Czuchry.
\newblock {Bounce scenarios in the Sotiriou-Visser-Weinfurtner generalization
  of the projectable Hořava-Lifshitz gravity}.
\newblock \emph{Class. Quant. Grav.}, 28:\penalty0 125013, 2011{\natexlab{a}}.
\newblock \doi{10.1088/0264-9381/28/12/125013}.

\bibitem[Czuchry(2011{\natexlab{b}})]{Czuchry:2009hz}
Ewa Czuchry.
\newblock {The Phase portrait of a matter bounce in Hořava-Lifshitz
  cosmology}.
\newblock \emph{Class. Quant. Grav.}, 28:\penalty0 085011, 2011{\natexlab{b}}.
\newblock \doi{10.1088/0264-9381/28/8/085011}.

\bibitem[Dutta and Saridakis(2010{\natexlab{a}})]{Dutta:2009jn}
Sourish Dutta and Emmanuel~N. Saridakis.
\newblock {Observational constraints on Hořava-Lifshitz cosmology}.
\newblock \emph{JCAP}, 1001:\penalty0 013, 2010{\natexlab{a}}.
\newblock \doi{10.1088/1475-7516/2010/01/013}.

\bibitem[Yagi et~al.(2014{\natexlab{a}})Yagi, Blas, Barausse, and
  Yunes]{Yagi:2013ava}
Kent Yagi, Diego Blas, Enrico Barausse, and Nicolás Yunes.
\newblock {Constraints on Einstein-Æther theory and Hořava gravity from
  binary pulsar observations}.
\newblock \emph{Phys. Rev.}, D89\penalty0 (8):\penalty0 084067,
  2014{\natexlab{a}}.
\newblock \doi{10.1103/PhysRevD.90.069902, 10.1103/PhysRevD.90.069901,
  10.1103/PhysRevD.89.084067}.
\newblock [Erratum: Phys. Rev.D90,no.6,069901(2014)].

\bibitem[Yagi et~al.(2014{\natexlab{b}})Yagi, Blas, Yunes, and
  Barausse]{Yagi:2013qpa}
Kent Yagi, Diego Blas, Nicolás Yunes, and Enrico Barausse.
\newblock {Strong Binary Pulsar Constraints on Lorentz Violation in Gravity}.
\newblock \emph{Phys. Rev. Lett.}, 112\penalty0 (16):\penalty0 161101,
  2014{\natexlab{b}}.
\newblock \doi{10.1103/PhysRevLett.112.161101}.

\bibitem[Park(2010)]{Park:2009zr}
Mu-in Park.
\newblock {A Test of Hořava Gravity: The Dark Energy}.
\newblock \emph{JCAP}, 1001:\penalty0 001, 2010.
\newblock \doi{10.1088/1475-7516/2010/01/001}.

\bibitem[Frusciante et~al.(2016)Frusciante, Raveri, Vernieri, Hu, and
  Silvestri]{Frusciante:2015maa}
Noemi Frusciante, Marco Raveri, Daniele Vernieri, Bin Hu, and Alessandra
  Silvestri.
\newblock {Hořava Gravity in the Effective Field Theory formalism: From
  cosmology to observational constraints}.
\newblock \emph{Phys. Dark Univ.}, 13:\penalty0 7--24, 2016.
\newblock \doi{10.1016/j.dark.2016.03.002}.

\bibitem[Blas et~al.(2010)Blas, Pujolas, and Sibiryakov]{Blas:2009qj}
D.~Blas, O.~Pujolas, and S.~Sibiryakov.
\newblock {Consistent Extension of Hořava Gravity}.
\newblock \emph{Phys. Rev. Lett.}, 104:\penalty0 181302, 2010.
\newblock \doi{10.1103/PhysRevLett.104.181302}.

\bibitem[Audren et~al.(2015)Audren, Blas, Ivanov, Lesgourgues, and
  Sibiryakov]{Audren:2014hza}
B.~Audren, D.~Blas, M.~M. Ivanov, J.~Lesgourgues, and S.~Sibiryakov.
\newblock {Cosmological constraints on deviations from Lorentz invariance in
  gravity and dark matter}.
\newblock \emph{JCAP}, 1503\penalty0 (03):\penalty0 016, 2015.
\newblock \doi{10.1088/1475-7516/2015/03/016}.

\bibitem[Audren et~al.(2013)Audren, Blas, Lesgourgues, and
  Sibiryakov]{Audren:2013dwa}
B.~Audren, D.~Blas, J.~Lesgourgues, and S.~Sibiryakov.
\newblock {Cosmological constraints on Lorentz violating dark energy}.
\newblock \emph{JCAP}, 1308:\penalty0 039, 2013.
\newblock \doi{10.1088/1475-7516/2013/08/039}.

\bibitem[Coleman and Glashow(1999)]{Coleman:1998ti}
Sidney~R. Coleman and Sheldon~L. Glashow.
\newblock {High-energy tests of Lorentz invariance}.
\newblock \emph{Phys. Rev.}, D59:\penalty0 116008, 1999.
\newblock \doi{10.1103/PhysRevD.59.116008}.

\bibitem[Kostelecky and Russell(2011)]{Kostelecky:2008ts}
V.~Alan Kostelecky and Neil Russell.
\newblock {Data Tables for Lorentz and CPT Violation}.
\newblock \emph{Rev. Mod. Phys.}, 83:\penalty0 11--31, 2011.
\newblock \doi{10.1103/RevModPhys.83.11}.

\bibitem[Wang(2017)]{Wang:2017brl}
Anzhong Wang.
\newblock {Hořava gravity at a Lifshitz point: A progress report}.
\newblock \emph{Int. J. Mod. Phys.}, D26\penalty0 (07):\penalty0 1730014, 2017.
\newblock \doi{10.1142/S0218271817300142}.

\bibitem[Sotiriou et~al.(2009)Sotiriou, Visser, and
  Weinfurtner]{Sotiriou:2009bx}
Thomas~P. Sotiriou, Matt Visser, and Silke Weinfurtner.
\newblock {Quantum gravity without Lorentz invariance}.
\newblock \emph{JHEP}, 10:\penalty0 033, 2009.
\newblock \doi{10.1088/1126-6708/2009/10/033}.

\bibitem[Sotiriou(2011)]{Sotiriou:2010wn}
Thomas~P. Sotiriou.
\newblock {Hořava-Lifshitz gravity: a status report}.
\newblock \emph{J. Phys. Conf. Ser.}, 283:\penalty0 012034, 2011.
\newblock \doi{10.1088/1742-6596/283/1/012034}.

\bibitem[Blas et~al.(2009)Blas, Pujolas, and Sibiryakov]{Blas:2009yd}
D.~Blas, O.~Pujolas, and S.~Sibiryakov.
\newblock {On the Extra Mode and Inconsistency of Horava Gravity}.
\newblock \emph{JHEP}, 10:\penalty0 029, 2009.
\newblock \doi{10.1088/1126-6708/2009/10/029}.

\bibitem[Charmousis et~al.(2009)Charmousis, Niz, Padilla, and
  Saffin]{Charmousis:2009tc}
Christos Charmousis, Gustavo Niz, Antonio Padilla, and Paul~M. Saffin.
\newblock {Strong coupling in Hořava gravity}.
\newblock \emph{JHEP}, 08:\penalty0 070, 2009.
\newblock \doi{10.1088/1126-6708/2009/08/070}.

\bibitem[Vernieri and Sotiriou(2012)]{Vernieri:2011aa}
Daniele Vernieri and Thomas~P. Sotiriou.
\newblock {Hořava-Lifshitz Gravity: Detailed Balance Revisited}.
\newblock \emph{Phys. Rev.}, D85:\penalty0 064003, 2012.
\newblock \doi{10.1103/PhysRevD.85.069901, 10.1103/PhysRevD.85.064003}.

\bibitem[Appignani et~al.(2010)Appignani, Casadio, and
  Shankaranarayanan]{Appignani:2009dy}
Corrado Appignani, Roberto Casadio, and S.~Shankaranarayanan.
\newblock {The Cosmological Constant and Horava-Lifshitz Gravity}.
\newblock \emph{JCAP}, 1004:\penalty0 006, 2010.
\newblock \doi{10.1088/1475-7516/2010/04/006}.

\bibitem[Vernieri(2015)]{Vernieri:2015uma}
Daniele Vernieri.
\newblock {On power-counting renormalizability of Hořava gravity with detailed
  balance}.
\newblock \emph{Phys. Rev.}, D91\penalty0 (12):\penalty0 124029, 2015.
\newblock \doi{10.1103/PhysRevD.91.124029}.

\bibitem[Colombo et~al.(2015)Colombo, Gümrükçüoğlu, and
  Sotiriou]{Colombo:2015yha}
Mattia Colombo, A.~Emir Gümrükçüoğlu, and Thomas~P. Sotiriou.
\newblock {Hořava gravity with mixed derivative terms: Power counting
  renormalizability with lower order dispersions}.
\newblock \emph{Phys. Rev.}, D92\penalty0 (6):\penalty0 064037, 2015.
\newblock \doi{10.1103/PhysRevD.92.064037}.

\bibitem[{L{\"u}} et~al.(2009){L{\"u}}, {Mei}, and {Pope}]{Lu:2009em}
H.~{L{\"u}}, J.~{Mei}, and C.~N. {Pope}.
\newblock Solutions to ho{\v r}ava gravity.
\newblock \emph{Phys. Rev. Lett.}, 103:\penalty0 091301, 2009.
\newblock \doi{10.1103/PhysRevLett.103.091301}.

\bibitem[Pospelov and Shang(2012)]{Pospelov:2010mp}
Maxim Pospelov and Yanwen Shang.
\newblock {On Lorentz violation in Horava-Lifshitz type theories}.
\newblock \emph{Phys. Rev.}, D85:\penalty0 105001, 2012.
\newblock \doi{10.1103/PhysRevD.85.105001}.

\bibitem[Coates et~al.(2016)Coates, Colombo, Gumrukcuoglu, and
  Sotiriou]{Coates:2016zvg}
Andrew Coates, Mattia Colombo, A.~Emir Gumrukcuoglu, and Thomas~P. Sotiriou.
\newblock {Uninvited guest in mixed derivative Hořava gravity}.
\newblock \emph{Phys. Rev.}, D94\penalty0 (8):\penalty0 084014, 2016.
\newblock \doi{10.1103/PhysRevD.94.084014}.

\bibitem[Carroll(2001)]{Carroll2001}
Sean~M. Carroll.
\newblock The cosmological constant.
\newblock \emph{Living Reviews in Relativity}, 4\penalty0 (1):\penalty0 1, Feb
  2001.
\newblock ISSN 1433-8351.
\newblock \doi{10.12942/lrr-2001-1}.
\newblock URL \url{https://doi.org/10.12942/lrr-2001-1}.

\bibitem[Scolnic et~al.(2017)]{Scolnic:2017caz}
D.~M. Scolnic et~al.
\newblock {The Complete Light-curve Sample of Spectroscopically Confirmed Type
  Ia Supernovae from Pan-STARRS1 and Cosmological Constraints from The Combined
  Pantheon Sample}.
\newblock 2017.
\newblock \doi{10.17909/T95Q4X}.

\bibitem[Liu and Wei(2015)]{Liu:2014vda}
Jing Liu and Hao Wei.
\newblock {Cosmological models and gamma-ray bursts calibrated by using Padé
  method}.
\newblock \emph{Gen. Rel. Grav.}, 47\penalty0 (11):\penalty0 141, 2015.
\newblock \doi{10.1007/s10714-015-1986-1}.

\bibitem[Ade et~al.(2016)]{planck2015}
P.~A.~R. Ade et~al.
\newblock {Planck 2015 results. XIII. Cosmological parameters}.
\newblock \emph{Astron. Astrophys.}, 594:\penalty0 A13, 2016.
\newblock \doi{10.1051/0004-6361/201525830}.

\bibitem[Moresco(2015)]{Moresco15}
Michele Moresco.
\newblock {Raising the bar: new constraints on the Hubble parameter with cosmic
  chronometers at z $\sim$ 2}.
\newblock \emph{Mon. Not. Roy. Astron. Soc.}, 450\penalty0 (1):\penalty0
  L16--L20, 2015.
\newblock \doi{10.1093/mnrasl/slv037}.

\bibitem[Betoule et~al.(2014)]{JLA}
M.~Betoule et~al.
\newblock {Improved cosmological constraints from a joint analysis of the
  SDSS-II and SNLS supernova samples}.
\newblock \emph{Astron. Astrophys.}, 568:\penalty0 A22, 2014.
\newblock \doi{10.1051/0004-6361/201423413}.

\bibitem[Blake et~al.(2012)]{WiggleZ}
Chris Blake et~al.
\newblock {The WiggleZ Dark Energy Survey: Joint measurements of the expansion
  and growth history at z < 1}.
\newblock \emph{Mon. Not. Roy. Astron. Soc.}, 425:\penalty0 405--414, 2012.
\newblock \doi{10.1111/j.1365-2966.2012.21473.x}.

\bibitem[Alam et~al.(2016)]{SDSS12}
Shadab Alam et~al.
\newblock {The clustering of galaxies in the completed SDSS-III Baryon
  Oscillation Spectroscopic Survey: cosmological analysis of the DR12 galaxy
  sample}.
\newblock \emph{Submitted to: Mon. Not. Roy. Astron. Soc.}, 2016.

\bibitem[Font-Ribera et~al.(2014)]{Lyman}
Andreu Font-Ribera et~al.
\newblock {Quasar-Lyman $\alpha$ Forest Cross-Correlation from BOSS DR11:
  Baryon Acoustic Oscillations}.
\newblock \emph{JCAP}, 1405:\penalty0 027, 2014.
\newblock \doi{10.1088/1475-7516/2014/05/027}.

\bibitem[Bennett et~al.(2014)Bennett, Larson, Weiland, and Hinshaw]{Bennett14}
C.~L. Bennett, D.~Larson, J.~L. Weiland, and G.~Hinshaw.
\newblock {The 1\% Concordance Hubble Constant}.
\newblock \emph{Astrophys. J.}, 794:\penalty0 135, 2014.
\newblock \doi{10.1088/0004-637X/794/2/135}.

\bibitem[Bogdanos and Saridakis(2010)]{0264-9381-27-7-075005}
Charalampos Bogdanos and Emmanuel~N Saridakis.
\newblock Perturbative instabilities in hořava gravity.
\newblock \emph{Classical and Quantum Gravity}, 27\penalty0 (7):\penalty0
  075005, 2010.
\newblock URL \url{http://stacks.iop.org/0264-9381/27/i=7/a=075005}.

\bibitem[Olive et~al.(2000)Olive, Steigman, and Walker]{Olive:1999ij}
Keith~A. Olive, Gary Steigman, and Terry~P. Walker.
\newblock {Primordial nucleosynthesis: Theory and observations}.
\newblock \emph{Phys. Rept.}, 333:\penalty0 389--407, 2000.
\newblock \doi{10.1016/S0370-1573(00)00031-4}.

\bibitem[Hagiwara et~al.(2002)]{Hagiwara:2002fs}
Kaoru Hagiwara et~al.
\newblock {Review of particle physics. Particle Data Group}.
\newblock \emph{Phys. Rev.}, D66:\penalty0 010001, 2002.
\newblock \doi{10.1103/PhysRevD.66.010001}.

\bibitem[Steigman(2006)]{Steigman:2005uz}
Gary Steigman.
\newblock {Primordial nucleosynthesis: successes and challenges}.
\newblock \emph{Int. J. Mod. Phys.}, E15:\penalty0 1--36, 2006.
\newblock \doi{10.1142/S0218301306004028}.

\bibitem[Cyburt et~al.(2016)Cyburt, Fields, Olive, and Yeh]{Cyburt:2015mya}
Richard~H. Cyburt, Brian~D. Fields, Keith~A. Olive, and Tsung-Han Yeh.
\newblock {Big Bang Nucleosynthesis: 2015}.
\newblock \emph{Rev. Mod. Phys.}, 88:\penalty0 015004, 2016.
\newblock \doi{10.1103/RevModPhys.88.015004}.

\bibitem[Oldengott and Schwarz(2017)]{Oldengott:2017tzj}
Isabel~M. Oldengott and Dominik~J. Schwarz.
\newblock {Improved constraints on lepton asymmetry from the cosmic microwave
  background}.
\newblock \emph{EPL}, 119\penalty0 (2):\penalty0 29001, 2017.
\newblock \doi{10.1209/0295-5075/119/29001}.

\bibitem[Zentner and Walker(2002)]{Zentner:2001zr}
Andrew~R. Zentner and Terry~P. Walker.
\newblock {Constraints on the cosmological relativistic energy density}.
\newblock \emph{Phys. Rev.}, D65:\penalty0 063506, 2002.
\newblock \doi{10.1103/PhysRevD.65.063506}.

\bibitem[Bean et~al.(2002)Bean, Hansen, and Melchiorri]{Bean:2002sm}
Rachel Bean, Steen~H. Hansen, and Alessandro Melchiorri.
\newblock {Constraining the dark universe}.
\newblock \emph{Nucl. Phys. Proc. Suppl.}, 110:\penalty0 167--172, 2002.
\newblock \doi{10.1016/S0920-5632(02)01475-5}.

\bibitem[{Bonvin} et~al.(2017){Bonvin}, {Courbin}, {Suyu}, {Marshall}, {Rusu},
  {Sluse}, {Tewes}, {Wong}, {Collett}, {Fassnacht}, {Treu}, {Auger}, {Hilbert},
  {Koopmans}, {Meylan}, {Rumbaugh}, {Sonnenfeld}, and {Spiniello}]{Bonvin}
V.~{Bonvin}, F.~{Courbin}, S.~H. {Suyu}, P.~J. {Marshall}, C.~E. {Rusu},
  D.~{Sluse}, M.~{Tewes}, K.~C. {Wong}, T.~{Collett}, C.~D. {Fassnacht},
  T.~{Treu}, M.~W. {Auger}, S.~{Hilbert}, L.~V.~E. {Koopmans}, G.~{Meylan},
  N.~{Rumbaugh}, A.~{Sonnenfeld}, and C.~{Spiniello}.
\newblock {H0LiCOW - V. New COSMOGRAIL time delays of HE 0435-1223: H$_{0}$ to
  3.8 per cent precision from strong lensing in a flat {$\Lambda$}CDM model}.
\newblock \emph{MNRAS}, 465:\penalty0 4914--4930, March 2017.
\newblock \doi{10.1093/mnras/stw3006}.

\bibitem[{Riess} et~al.(2011){Riess}, {Macri}, {Casertano}, {Lampeitl},
  {Ferguson}, {Filippenko}, {Jha}, {Li}, and {Chornock}]{Riess}
A.~G. {Riess}, L.~{Macri}, S.~{Casertano}, H.~{Lampeitl}, H.~C. {Ferguson},
  A.~V. {Filippenko}, S.~W. {Jha}, W.~{Li}, and R.~{Chornock}.
\newblock {A 3\% Solution: Determination of the Hubble Constant with the Hubble
  Space Telescope and Wide Field Camera 3}.
\newblock \emph{ApJ}, 730:\penalty0 119, April 2011.
\newblock \doi{10.1088/0004-637X/730/2/119}.

\bibitem[Dutta and Saridakis(2010{\natexlab{b}})]{Dutta:2010jh}
Sourish Dutta and Emmanuel~N. Saridakis.
\newblock {Overall observational constraints on the running parameter $\lambda$
  of Horava-Lifshitz gravity}.
\newblock \emph{JCAP}, 1005:\penalty0 013, 2010{\natexlab{b}}.
\newblock \doi{10.1088/1475-7516/2010/05/013}.

\bibitem[Nilsson and Dabrowski(2017)]{Nilsson:2016rsv}
Nils~Albin Nilsson and Mariusz~P. Dabrowski.
\newblock {Energy Scale of Lorentz Violation in Rainbow Gravity}.
\newblock \emph{Phys. Dark Univ.}, 18:\penalty0 115--122, 2017.
\newblock \doi{10.1016/j.dark.2017.10.003}.

\bibitem[Salzano and Dabrowski(2017)]{Salzano:2016pny}
Vincenzo Salzano and Mariusz~P. Dabrowski.
\newblock {Statistical hierarchy of varying speed of light cosmologies}.
\newblock \emph{Astrophys. J.}, 851\penalty0 (2):\penalty0 97, 2017.
\newblock \doi{10.3847/1538-4357/aa9cea}.

\bibitem[Lewis and Bridle(2002)]{Lewis:2002ah}
Antony Lewis and Sarah Bridle.
\newblock {Cosmological parameters from CMB and other data: A Monte Carlo
  approach}.
\newblock \emph{Phys. Rev.}, D66:\penalty0 103511, 2002.
\newblock \doi{10.1103/PhysRevD.66.103511}.

\bibitem[{Audren} et~al.(2013){Audren}, {Lesgourgues}, {Benabed}, and
  {Prunet}]{2013JCAP...02..001A}
B.~{Audren}, J.~{Lesgourgues}, K.~{Benabed}, and S.~{Prunet}.
\newblock {Conservative constraints on early cosmology with MONTE PYTHON}.
\newblock \emph{JCAP}, 2:\penalty0 001, February 2013.
\newblock \doi{10.1088/1475-7516/2013/02/001}.

\bibitem[Hogg(1999)]{Hogg}
David~W. Hogg.
\newblock {Distance measures in cosmology}.
\newblock 1999.

\bibitem[Eisenstein and Hu(1998)]{Eisenstein}
Daniel~J. Eisenstein and Wayne Hu.
\newblock {Baryonic features in the matter transfer function}.
\newblock \emph{Astrophys. J.}, 496:\penalty0 605, 1998.
\newblock \doi{10.1086/305424}.

\bibitem[Wang and Dai(2016)]{WangWang}
Yun Wang and Mi~Dai.
\newblock {Exploring uncertainties in dark energy constraints using current
  observational data with Planck 2015 distance priors}.
\newblock \emph{Phys. Rev.}, D94\penalty0 (8):\penalty0 083521, 2016.
\newblock \doi{10.1103/PhysRevD.94.083521}.

\bibitem[Hu and Sugiyama(1996)]{Hu}
Wayne Hu and Naoshi Sugiyama.
\newblock {Small scale cosmological perturbations: An Analytic approach}.
\newblock \emph{Astrophys. J.}, 471:\penalty0 542--570, 1996.
\newblock \doi{10.1086/177989}.

\end{thebibliography}
\end{document}